\documentclass[aps,superscriptaddress,eqsecnum,nofootinbib,showpacs,preprintnumbers]{revtex4-2}
\usepackage[utf8]{inputenc}
\usepackage{hyperref}
\usepackage{amsmath}
\usepackage{mathtools}
\usepackage{amsfonts}
\usepackage{latexsym}
\usepackage{xcolor}
\usepackage{graphicx}
\usepackage{float}
\usepackage{xcolor}
\usepackage{framed}
\usepackage{amssymb}
\usepackage{graphicx, rotating}
\usepackage{epsfig}
\usepackage{latexsym}
\usepackage{amsmath,bm,amssymb}
\usepackage{slashed}
\usepackage{soul}
\usepackage{hyperref}
\usepackage{comment}
\usepackage{subcaption}
\usepackage{epstopdf}
\usepackage{lipsum}  %random text
\usepackage{slashed}
\usepackage{upgreek} %extra greek letters
\newcommand*\diff{\mathop{}\!\mathrm{d}}

\graphicspath{{Figs/}}

\definecolor{azure(colorwheel)}{rgb}{0.0, 0.5, 1.0}
\definecolor{MyDarkBlue}{rgb}{0,0.1,0.7}
\definecolor{DarkViolet}{RGB}{148,0,211}
\definecolor{DarkBlue}{RGB}{0,0,153}
\definecolor{amber}{rgb}{1.0, 0.49, 0.0}
\definecolor{amaranth}{rgb}{0.9, 0.17, 0.31}
\definecolor{nicered}{rgb}{0.7,0.1,0.1}
\definecolor{brown}{rgb}{0.5,0.1,0.1}
\definecolor{nicegreen}{rgb}{0.0,0.3,0.0}
\definecolor{tealgreen}{rgb}{0.0, 0.51, 0.5}
\hypersetup{colorlinks,bookmarksopen,
bookmarksnumbered,
citecolor={nicered},
linkcolor={MyDarkBlue},
urlcolor={tealgreen},
pdfstartview=FitH}

\newcommand{\newc}{\newcommand}
\newc{\teo}[1]{\textcolor{azure(colorwheel)}{#1}} % teo color
\newc{\com}[1]{\textcolor{amaranth}{#1}} % comment color
\newc{\bako}[1]{\textcolor{DarkViolet}{#1}} % bakop color
\newc{\corr}[1]{\textcolor{red}{#1}} % important color
\newc{\kar}[1]{\textcolor{DarkBlue}{#1}} % important color
\newc{\nick}[1]{\textcolor{magenta}{#1}} % important color

\usepackage{times}

\usepackage{orcidlink}
\def\idbako{\orcidlink{0000-0002-3012-6144}}
\def\idkara{\orcidlink{0000-0002-5479-6513}}
\def\idnakas{\orcidlink{0000-0002-3522-5803}}
\def\idpapa{\orcidlink{0000-0003-1244-922X}}
\def\idmavr{\orcidlink{0000-0001-7683-8625}}

\begin{document}
\preprint{\leftline{KCL-TH-PH/2024-{\bf 03}}}

\title{ Exact black holes in string-inspired Euler-Heisenberg theory}

	\author{Athanasios Bakopoulos\idbako}
	\email{atbakopoulos@gmail.com}

\affiliation{Physics Division, School of Applied Mathematical and Physical Sciences, National Technical University of Athens, Zografou Campus, Athens 15780, Greece}
 
	%\affiliation{Division of Applied Analysis, Department of Mathematics, University of Patras, Rio Patras GR-26504, Greece}

	\author{Thanasis Karakasis\idkara}
	\email{thanasiskarakasis@mail.ntua.gr}
	\affiliation{Physics Division, School of Applied Mathematical and Physical Sciences, National Technical University of Athens, Zografou Campus, Athens 15780, Greece}

    \author{Nick~E.~Mavromatos\idmavr}
    \email{mavroman@mail.ntua.gr} 

\affiliation{Physics Division, School of Applied Mathematical and Physical Sciences, National Technical University of Athens, Zografou Campus, Athens 15780, Greece}

    \affiliation{Theoretical Particle Physics and Cosmology group, Department of Physics, King's College London, London WC2R 2LS, UK.}

    \author{Theodoros Nakas\idnakas}
	\email{theodoros.nakas@gmail.com} 

\affiliation{Physics Division, School of Applied Mathematical and Physical Sciences, National Technical University of Athens,  Zografou Campus, Athens 15780, Greece}

	\author{Eleftherios Papantonopoulos\idpapa}
	\email{lpapa@central.ntua.gr} 

\affiliation{Physics Division, School of Applied Mathematical and Physical Sciences, National Technical University of Athens, Zografou Campus, Athens 15780, Greece}

\begin{abstract}
    We consider higher-order derivative gauge field corrections that arise in the fundamental context of dimensional reduction of String Theory and Lovelock-inspired gravities and obtain an exact and asymptotically flat black-hole solution, in the presence of non-trivial dilaton configurations. Specifically, by considering the gravitational theory of Euler-Heisenberg non-linear electrodynamics coupled to a dilaton field with specific coupling functions, we perform an extensive analysis of the characteristics of the black hole, including its geodesics for massive particles, the energy conditions, thermodynamical and stability analysis. The inclusion of a dilaton scalar potential in the action can also give rise to asymptotically (A)dS spacetimes and an effective cosmological constant. Moreover, we find that the black hole can be thermodynamically favored when compared to the Gibbons-Maeda-Garfinkle-Horowitz-Strominger (GMGHS) black hole for those parameters of the model that lead to a larger black-hole horizon for the same mass. Finally, it is observed that the energy conditions of the obtained black hole are indeed satisfied, further validating the robustness of the solution within the theoretical framework, but also implying that this self-gravitating dilaton-non-linear-electrodynamics system constitutes another explicit example of bypassing modern versions of the no-hair theorem without any violation of the energy conditions.
\end{abstract}
\maketitle
\tableofcontents

\section{Introduction}\label{sec:intro}

In the pursuit of a comprehensive understanding of gravitational phenomena in the cosmos as well as gravity itself, the theoretical examination of black holes stands as an essential frontier. 
The General theory of Relativity (GR), while highly successful in describing the macroscopic behavior of these celestial entities, becomes subject to scrutiny under extreme conditions. 
This investigation prompts the exploration of modified gravitational theories and theories with extra dimensions.
Among the theories attempting to unify fundamental interactions, String Theory stands as the leading contender.
In particular, the heterotic string theory stands as an essential branch within the broader scope of String Theory, distinguished by its capability to unify gravitational interactions with other fundamental forces. 
Notably, it excels in synthesizing these interactions into a cohesive framework. 
A focal point of interest lies in the derivation of an effective four-dimensional theory, offering insights into quantum corrections that modify Einstein's theory of gravity.
These corrections can potentially incorporate terms ranging from the Gauss-Bonnet, quadratic-curvature term~\cite{Gross:1986mw,Green:2012oqa,Green:2012pqa,Kanti:1995vq} to non-linear electromagnetic corrections (see e.g. \cite{Natsuume:1994hd, Kats:2006xp, Anninos:2008sj, Liu:2008kt} and references within).

Drawing therefore inspiration only from the aforementioned corrections introduced by string/brane theory, in this article, we aim to elucidate the implications of departing from the conventional electromagnetic framework and embracing the intricacies of non-linear electrodynamics (NED) and scalar fields within the context of black hole solutions.
In addition to the above, it is important to mention that four-dimensional scalar-vector-tensor theories can be also obtained via an appropriate reduction from a higher-dimensional Lovelock theory \cite{Charmousis:2012dw}. Under this perspective, scalar-tensor-vector theories can be understood as natural extensions of the scalar-tensor theories that have been extensively studied in the last decades \cite{Kanti:1995vq, Babichev:2013cya,Babichev:2016rlq,Babichev:2017guv,Charmousis:2011bf,Herdeiro:2015waa,Baake:2023zsq, Antoniou:2017acq,Antoniou:2017hxj,Antoniou:2019awm,Bakopoulos:2018nui,Bakopoulos:2019tvc,Bakopoulos:2020dfg,Bakopoulos:2021dry,Bakopoulos:2021liw,Bakopoulos:2022csr,Bakopoulos:2023fmv,Bakopoulos:2023hkh,Bakopoulos:2023sdm,Bakopoulos:2023tso,Nakas:2023yhj,Chatzifotis:2021hpg,Chatzifotis:2022mob,Chatzifotis:2022ubq,Karakasis:2021lnq,Karakasis:2021rpn,Karakasis:2021tqx,Karakasis:2021ttn,Karakasis:2023hni,Karakasis:2023ljt,Tang:2019jkn,Tang:2020sjs,Liu:2020yqa,Babichev:2022awg,Babichev:2023dhs,Babichev:2023mgk,Babichev:2023psy,Babichev:2023rhn,Charmousis:2021npl,Antoniou:2021zoy,Andreou:2019ikc,Ventagli:2020rnx,Mahapatra:2020wym, Chew:2024rin, Chew:2023olq, Chew:2022enh}.  
Such scalar-vector-tensor theories offer a very fruitful framework for finding novel compact-object solutions, eluding the constraints imposed by the ``no-hair" theorems \cite{Bekenstein:1971, Bekenstein:1972ny, Teitelboim:1972, PhysRevD.51.R6608, Mayo:1996mv, Mazur:2000pn, Chrusciel:2012jk,Priyadarshinee:2021rch,Priyadarshinee:2023cmi,Daripa:2024ksg,Theodosopoulos:2023ice,Karakasis:2022fep,Karakasis:2022xzm,KordZangeneh:2017zgg,Tang:2016vmu,Sanchis-Gual:2022mkk,Herdeiro:2016tmi,Herdeiro:2018wub,Charmousis:2009xr,Babichev:2017rti,Babichev:2015rva,Barrientos:2022bzm,Cisterna:2016nwq,Cisterna:2020rkc,Cisterna:2021ckn,Rehman:2023eor,Rehman:2023hro,Magos:2023nnb,Dernek:2023zbx,Ali:2023kdp}. 
This departure from traditional limitations is attributed to the dilaton field, which, notably, introduces no additional independent free parameter into the resulting solution \cite{Kiorpelidi:2023jjw}. 
Instead, these solutions manifest a distinct feature known as ``secondary hair", intricately determined by the compact object's mass, charge, and angular momentum.

Despite the motivation coming from higher-dimensional theories, there are also additional reasons that lead us to explore non-linear electrodynamics.
First and foremost, in regions with strong gravitational fields, such as those near black holes, traditional linear theories may break down. 
Non-linear electrodynamics becomes important in these strong field regimes, where the intensity of electromagnetic fields can become comparable to the strength of gravitational fields. 
Studying how non-linearities affect the behavior of electromagnetic fields in these regimes is crucial for understanding the physics of objects like black holes, neutron stars, and other astrophysical phenomena.
Moreover, non-linear electrodynamics is expected to lead to phenomena that are absent in linear theories. 
In the early universe for example, when energy densities were extremely high, the interplay between gravitational and electromagnetic fields was significant. 
NED can be crucial in modeling the behavior of these fields during cosmological evolution. 
Consequently, NED allows us to investigate how electromagnetic interactions influenced the dynamics of the early universe and whether non-linear effects played a role in the formation of cosmic structures. 
Understanding these cosmological implications helps build a more complete picture of the evolution of the universe.
For a review on non-linear electrodynamics and its applications, see \cite{Sorokin:2021tge} and references therein.

In this paper, we embark on a comprehensive exploration of a gravitational theory extending the classical Euler-Heisenberg (EH) electrodynamics coupled to a non-trivial dilaton field. Our motivation for this study stems from the rich theoretical landscape it promises, building upon the established framework of self-gravitating dilaton-linear-electrodynamics. This extension allows us to delve into intriguing phenomena, notably exemplified by the Gibbons-Maeda-Garfinkle-Horowitz-Strominger (GMGHS) black hole~\cite{Gibbons:1987ps, Garfinkle:1990qj}, a significant exact solution within this domain. In our investigation, we examine the intricacies of our proposed model and unravel its associated black-hole solution in detail. One of our key insights lies in the strategic assumption of a specific profile governing the dilaton coupling to the Euler-Heisenberg terms. 
This choice results in an exact analytic black-hole solution, facilitating a straightforward examination of its physical characteristics.

Having the solution at hand, we then commence a rigorous analysis encompassing various facets of our model's implications. This includes a thorough examination of the geodesics of massive test particles within the black-hole spacetime, followed by a meticulous scrutiny of the energy conditions. Subsequently, we delve into the thermodynamic aspects of the black hole, computing the relevant thermodynamic quantities, such as the temperature, the entropy, and the magnetic potential ($\Phi_m$), to demonstrate the validity of the first law of thermodynamics. Moreover, within the parameter space of solutions, we unveil the existence of pairs consisting of two distinct black holes characterized by different ratios $Q_m/M$, both more compact than the Schwarzschild solution yet sharing identical horizon radii. 
Intriguingly, despite their geometric similarity, a thermodynamic analysis reveals clear distinctions, with one black hole exhibiting thermodynamic stability while its \emph{doppelg\"{a}nger} proves to be thermodynamically unstable.
Additionally, we explore the radial stability of the black-hole solution under linear perturbations and also its scalar quasi-normal modes, shedding light on its potential as an astrophysical entity. 
Furthermore, we extend our discussions to encompass other solutions and extensions of our model theory, including asymptotically (Anti-)de Sitter (AdS) spacetimes and more general dilaton couplings, providing a comprehensive overview of the theoretical landscape.
In conclusion, our work offers a thorough investigation into the gravitational theory of non-linear EH electrodynamics coupled to a non-trivial dilaton field, unraveling a plethora of intriguing phenomena and paving the way for further exploration and theoretical advancements in this domain.

The structure of the current article is the following: In the next section \ref{sec:ned} we motivate our study by discussing a gravitational theory of non-linear Euler-Heisenberg (EH) electrodynamics coupled to a non-trivial dilaton field. This generalises the corresponding self-gravitating dilaton-linear-electrodynamics
case, known to  
admit the GMGHS black hole \cite{Gibbons:1987ps, Garfinkle:1990qj} as an exact solution. In the subsequent section \ref{sec:model}, we discuss our model and its associated black hole solution. By assuming a specific profile for the dilaton coupling to the EH terms, in such a way that, additionally to non-trivial dilaton couplings, one has also dilaton-independent EH terms, 
we demonstrate the possibility of studying analytically the corresponding black-hole solution. In section \ref{sec:geodener}
we first discuss the geodesics of test particles in such black hole spacetimes, and then demonstrate the satisfaction of the energy conditions for appropriate sets of the parameters of the solution. In section \ref{sec:thermo}, we study the thermodynamics of the black hole, and show explicitly, by computing the relevant thermodynamical quantities, that the first law of thermodynamics is satisfied in a coordinate-independent way, as should have been expected. 
In the parameter space of solutions, it is possible to obtain two distinct black holes with different ratios $Q_m/M$ that are more compact than the Schwarzschild solution and share the same horizon radius.
However, these black holes even though they have the same horizon radius, from a thermodynamic point of view, are quite distinguishable, since the solution with a greater value for the ratio $Q_m/M$ is thermodynamically stable, while its doppelg\"{a}nger with a lower value for the ratio $Q_m/M$ is thermodynamically unstable.
In section \ref{sec:linear}, we demonstrate the radial stability of the black-hole solution under linear perturbations, and study its scalar quasi-normal models, which provide insights into its properties as a potential astrophysical object. Other solutions of (extensions of) our model theory \eqref{theory}, including asymptotically (Anti-)de Sitter (AdS) spacetimes, as well as solutions corresponding to more general couplings $\exp(-2\gamma\phi)$, of the dilaton to the Maxwell term in the action, rather than the $\gamma=1$ in closed strings, 
are discussed in section \ref{sec:othersol}. Finally, conclusions and outlook are given in section \ref{sec:concl}. 

\section{String inspired non-linear electrodynamics}\label{sec:ned}

 One of the particular aspects of string/brane-induced non-linear electrodynamics effects is that the higher order in the Maxwell tensor can be combined into an all-order expression, the so-called Born-Infeld (BI) Lagrangian~\cite{Born:1933qff, Born:1934ji, Born-Infeld, Metsaev:1987qp, Andreev:1988cb, Tseytlin:1999dj}, as a result of re-summation of open string excitations (attached to, e.g., 3-brane worlds in the D-brane extension of string theory, in which case the world-volume of ($d=3$)-brane leads to the DiracBI (DBI) action~(see \cite{Leigh:1989jq,Dai:1989ua,Polchinski:1996na} and references therein). In such models, the BI electrodynamics in four spacetime dimensions originates from the higher ($d=10$)-dimensional superstring action upon either compactification or appropriate restriction on a $3d$-brane volume. It is important to note that in all such string-inspired models the BI Lagrangian couples to the inverse of the open-string coupling $g_s=e^{\phi}$, where $\phi$ is the (dimensionless) dilaton field, so the corresponding four-dimensional action in a curved four-dimensional background metric (in the Jordan or $\sigma$-model frame), $g^J_{\mu\nu}$, 
 %$\mu,\nu=0, \dots 3,$ 
 reads
\begin{align}\label{BI}
\mathcal S^J_{\rm BI} = - \mathcal T_4^2 \int d^4x \, e^{-\phi}\sqrt{{\rm Det}\Big(-g^J_{\mu\nu} + \mathcal T_4^{-1}\,    \mathcal F_{\mu\nu}\Big)} \,,
\end{align}
where $\mathcal{F}_{\mu\nu}$ is  the Maxwell tensor $\mathcal{F}_{\mu\nu} = \partial_{\mu}\mathcal{A}_{\nu} - \partial_{\nu}\mathcal{A}_{\mu}$, 
%Greek indices denote four-dimensional spacetime indices,
and $\mathcal T_4 =\frac{1}{2\pi\alpha^\prime} = \frac{M_s^2}{2\pi}\,,$
%\begin{align}\label{Msval}
%\mathcal T_4 =\frac{1}{2\pi\alpha^\prime} = \frac{M_s^2}{2\pi}\,,  
%\end{align}
is the (open) string tension, with $\alpha^\prime = M_s^{-2}$ the Regge slope ($M_s$ the string mass scale, which in general is different from the four-dimensional Planck scale). One may go away from string/brane theory and define the BI action as a starting point of an effective modified electrodynamics field theory. In such a case  the string tension 
$\mathcal T_4$ may be considered as an arbitrary phenomenological dimensionful parameter which is not related to the Regge slope $\alpha^\prime$. We term such a parameter the {\it BI parameter}.

We next remark that, in four spacetime dimensions, the determinant in the argument of the square root in the BI action \eqref{BI} can be expanded exactly to yield:
\begin{align}\label{ffdual}
 \mathcal S_{\rm BI}  = - \mathcal T_4^2 \int d^4x \, \sqrt{-g^J} \, e^{-\phi}\sqrt{1 + \frac{1}{2\, \mathcal T_4^2} \mathcal F_{\mu\nu}\, \mathcal F^{\mu\nu} -\frac{1}{16\, \mathcal T_4^4}\,\Big(\mathcal F_{\mu\nu}\, \widetilde{\mathcal F}^{\mu\nu}\Big)^2 }\, 
\end{align}
where $\widetilde{ \mathcal F}_{\mu\nu} = \frac{1}{2} \varepsilon_{\mu\nu\rho\sigma}\, \mathcal F^{\rho\sigma}$ is the dual of the Maxwell tensor,
with $\varepsilon_{\mu\nu\rho\sigma}$ the Levi-Civita fully antisymmetric symbol in curved spacetime with metric $g^J_{\mu\nu}$. Expanding the (square root in the) four-dimensional BI action \eqref{ffdual} in inverse powers of the BI parameter $\mathcal T_4$, leads to effective dimension $8$ (and higher) operators in the effective field theory, which make contact with the generic Euler-Heisenberg (EH) non-linear electrodynamics (NED)~\cite{Tseytlin:1999dj,Ellis:2017edi,Ellis:2022uxv}:
\begin{align}\label{expBI}
 \mathcal S_{\rm BI}  &=  \int d^4x \sqrt{-g^J} \, e^{-\phi} \Big[- \mathcal T_4^2 \, I_2 - \mathcal T_4^4 \, I_4 \Big(1 + \mathcal O(\mathcal F^2)\Big)\Big]\, , \nonumber \\
 I_2 &= \frac{1}{4\mathcal T_4^2} \, \mathcal F_{\mu\nu}\, \mathcal F^{\mu\nu} \,, \quad 
 I_4 = -\frac{1}{8\, \mathcal T_4^4} \mathcal F_{\mu\nu}\, \mathcal F^{\nu\rho}\, \mathcal F_{\rho\lambda} \,  \mathcal F^{\lambda\mu} + \frac{1}{32\, \mathcal T_4^4} \, \Big(\mathcal F_{\mu\nu}\, \mathcal F^{\mu\nu}\Big)^2\,.
\end{align}
Hence, ignoring for the moment the dilaton, to fourth order in the field strength $\mathcal{F}_{\mu\nu}$ one obtains (up to the dilaton factors) a special case of the generic EH NED with dimension $8$ operators, with Lagrangian:
\begin{align}\label{EHLag}
\mathcal L_{\rm EH} =  c_1\, \Big(\mathcal F_{\mu\nu}\, \mathcal F^{\mu\nu}\Big)^2 + c_2  \, 
\mathcal F_{\mu\nu}\, \mathcal F^{\nu\rho}\, \mathcal F_{\rho\lambda} \, \mathcal F^{\lambda\mu} \,,
\end{align}
where the BI Lagrangian corresponds to~\cite{Ellis:2017edi,Ellis:2022uxv} 
\begin{align}\label{cis}
 c_1 = -\frac{1}{32\mathcal T_4^2}\,, \quad 
 c_2 = \frac{1}{8\, \mathcal T_4^2}\,.
\end{align}
The reader should notice that the ratio of $c_2/c_1=-4$ exactly, which is a characteristic prediction of the BI theory.

Phenomenologically, assuming a constant dilaton and flat Minkowski spacetime, the BI parameter $\mathcal T_4^2$ can be constrained in collider physics, via light-by-light scattering, for which there is clear experimental evidence these days at LHC experiments (see ~\cite{dEnterria:2013zqi,ATLAS:2017fur,CMS:2018erd}). Such light-by-light scattering studies~\cite{Ellis:2017edi} can place a lower bound on the BI parameter $\mathcal T_4 \gtrsim 100~{\rm GeV}$. In the case of string theory, this would 
lead to a (weak) lower bound of the string
mass scale $M_s \gtrsim 0.25~{\rm TeV}$. Notably, extra dimension collider (LHC) searches place currently this bound to $M_s \gtrsim \mathcal O(10)$~TeV. Forecasts for much larger values of the lower bounds of the BI parameter in future colliders, in particular FCC, have been given in \cite{Ellis:2022uxv}.
Embedding the BI (or more generally Euler-Heisenberg) theory into curved spacetime, and fully incorporating the dilaton effects, leads to a whole new area of tests of NED by employing the entire machinery of modern gravitational experiments technology.

The BI action $\mathcal S_{\rm BI}$ \eqref{BI} and \eqref{ffdual} in curved background metrics can be augmented, at an effective field theory level, by including the dynamics of the gravitational ($g^J_{\mu\nu}$) and dilaton ($\phi$) fields. In this respect, we recall that the D-brane action is by construction in the so-called Jordan (or $\sigma$-model) frame. Passing into the Einstein frame in four dimensions, via the transformation of the metric: $g^J_{\mu\nu} \to g_{\mu\nu} = e^{-2\phi} g^J_{\mu\nu}$, we write for the pertinent gravitational action (in geometrized units $c=G=1$, in which we work from now on):
\begin{align}\label{gravBI}
    \mathcal{S} =\frac{1}{16\pi} \int \diff^4x \sqrt{-g} \Big[\mathcal{R} - 2\nabla^{\mu}\phi\nabla_{\mu}\phi \Big] -\int \diff^4x \, \sqrt{-g}\, e^{-\phi} \Big[\mathcal T_4^2\, I_2^{\rm E} + \mathcal T_4^4 \, e^{-4\phi}\, I_4^{\rm E}\Big] + \dots   
    \,,
\end{align}
where the quantities $I_i^{\rm E},\, i=2,4$ are given by the corresponding ones in \eqref{expBI}, but the indices
contraction is made by the Einstein-frame metric $g_{\mu\nu}$. 

Departing from the case of the brane DBI action \eqref{BI}, one may consider higher-order (in derivatives, that is in a Regge slope $\alpha^\prime$ expression) electromagnetic terms in effective low energy field theories stemming only from closed strings, e.g. the heterotic string~\cite{Gross:1986mw}. In such theories, unlike the DBI brane or open-string case, there is no resummation in closed form of the gauge terms. Nonetheless, some authors have 
generalised the BI effective action in a curved $(3+1)$-dimensional spacetime, by considering the following form of dilaton couplings to the electromagnetic fields in a BI NED setting~\cite{Yazadjiev:2005za,Dehghani:2006zi,Sheykhi:2006dz}:
\begin{align}\label{theory2}
    \mathcal{S} &=\frac{1}{16\pi} \int \diff^4x \sqrt{-g} \Big[\mathcal{R} - 2\nabla^{\mu}\phi\nabla_{\mu}\phi + \mathcal L_{\rm BI}\Big]~, \nonumber \\
    \mathcal L_{\rm BI} &= 4\beta_{\rm BI} \, e^{2\gamma \phi} \, 
    \Big(1- \sqrt{ 1 + \frac{e^{-4\gamma \phi}}{2\beta_{\rm BI}} \mathcal F^2 - 
    \frac{e^{-8\gamma \phi}}{16\, \beta_{\rm BI}^2} (\mathcal F \widetilde{\mathcal F})^2}\,\,\,\Big)
\end{align}
where the notation has been defined above, $\gamma$ defines the dilaton coupling, and now $\beta_{\rm BI}$ plays the role of the generalised BI parameter, with mass dimensions $+2$ (which is identified with $\mathcal T_4^2$ in the case of strings, in which case, to match with the corresponding $\mathcal O(\alpha^\prime)$ Maxwell terms of the heterotic-string effective action~\cite{Gross:1986mw}, 
$e^{-2\phi}\mathcal F^2$, one should fix $\gamma=1$). 

The above considerations deal with tree-level in string loops, that is first quantised actions on world-sheet with trivial topology ($2d$ sphere for closed string sectors, and disc for open one).
In general, string loop effective actions are not known in closed form. In simplified phenomenological scenarios such effective actions can be expressed in the generic form, e.g. in the closed string sector in the string (or $\sigma$-model frame with metric $\widehat g_{\mu\nu}$ in ($3+1$)-dimensions, after string compactification)~\cite{Damour:1994zq}:
\begin{align}\label{effactclsf}
S = \int \diff^4x \, \sqrt{-\widehat{g}}\, 
\Big(\frac{1}{\alpha^\prime} \, B_g(\Phi) \, \widehat{\mathcal R} + \frac{1}{\alpha^\prime}\, B_\Phi(\Phi) \, \Big[\widehat{\Box}\Phi - 4\widehat{\nabla}_\mu \Phi \, \widehat{\nabla}^\mu \Phi \Big] -\frac{B_F(\Phi)}{4}\widehat F_{\mu\nu}\, \widehat F^{\mu\nu} - B_\psi(\Phi)\overline{\widehat \psi}\, \slashed{\widehat D}\widehat \psi + \dots \Big)\,,
\end{align}
where the $\widehat{(...)} $ symbol above a tensorial quantity implies contraction of the world-indices with the string-frame metric $\widehat g_{\mu\nu}$, $F_{\mu\nu}$ denotes the field strength of the gauge field, $D_\mu$ is the gauge covariant derivative, $\psi$ are fermionic matter fields and the $\dots$ denote other matter fields as well as (an infinity of) higher-derivative (higher order in $\alpha^\prime$) terms, The quantities $B_i(\Phi), i=g,\Phi,F,\psi$ are non-derivative functions of the dilaton which arise from summing over (closed) world-sheet topologies, that is these functions involve powers of the string coupling $g_s=\exp(\Phi)$ of the form $g_s^{-\chi}$, where $\chi = 2- 2N$ where $N$ denotes the number of handles,  is the genus of the world-sheet surface (sphere has $N=0$, torus (one string loop) $=0$ etc. 
Thus, 
\begin{align}\label{tadpoles}
B_i(\Phi) = e^{-2\Phi} + c_0^{(i)} + c_1^{(i)} e^{2\Phi} + \dots + c_{2n}^{(i)}\, e^{2n\Phi} + \dots \,,
\end{align}
where the constant quantities $c_i$ pertain to effects of string loops, so that the expressions \eqref{tadpoles} involve a power series in the square of the string coupling $g_s^2=\exp(2\Phi)$.
The first term on the right-hand side of \eqref{tadpoles}
 leads to the standard closed string expression for the gauge field Maxwell terms in the low-energy effective action, $e^{-2\phi}\mathcal F^2$ for standard dilaton kinetic term normalization in the Einstein frame~\cite{Gross:1986mw,Green:2012oqa,Green:2012pqa}).\footnote{We note for completion that a similar factor accompanies the quadratic gravitational curvature (Gauss-Bonnet (GB)) terms in the action at string-loop tree level. This is a remnant of the corresponding situation of the ten-dimensional target-spacetime heterotic-string effective field theory action, which in the extra (compact) dimensional sector leads to the celebrated anomaly cancellation by equating the extra-dimensional (non-Abelian) gauge with the corresponding quadratic-curvature gravitational GB terms $\int d^6x \sqrt{-G^{(6)}} e^{-2\Phi} \Big({\rm Tr}\mathbf F^2 - \mathbf R_{\rm GB}^2\Big) \to 0$ (with the Tr being a group-index trace), which leads to the Heterotic string selecting the $E_8 \times E_8 $ gauge group as the unique target-space group before compactification to (3+1)-dimensions~\cite{Green:2012oqa,Green:2012pqa}.}

Passing to the Einstein frame, via appropriate redefinitions of~\cite{Damour:1994zq}: the metric $\widehat g_{\mu\nu} \rightarrow g_{\mu\nu} = C \, B_g(\Phi) \widehat g_{\mu\nu}$, where $C$ are numerical normalization constants, the dilaton $\Phi \rightarrow \phi = \int \diff\Phi\, \sqrt{\frac{3}{4}\Big(\frac{B_g^\prime}{B_g}\Big)^2 + 2(\frac{B_\Phi^\prime}{B_\Phi} + \frac{B_\Phi}{B_g})}$, where the prime denotes $d/d\Phi$, and the fermionic matter fields, $\widehat \psi \rightarrow \psi = C^{-3/4}\, B_g^{-3/4}\, B_\psi^{1/2}\, \widehat \psi$, leaving the gauge fields as they are, yields the effective action: 
\begin{align}\label{effactcl}
S&= \int d^4 x \sqrt{-g} \frac{1}{8\pi\overline{\rm G}} \left(\mathcal{R}- 2\nabla_\mu \phi \nabla^\mu \phi\right) + 
S_{\rm matter}\,, \nonumber \\
S_{\rm matter} &= \int \diff^4x\, \sqrt{-g} \Big[-\overline\psi \slashed{D} \psi - \frac{1}{4}B_F(\Phi)F_{\mu\nu}\, F^{\mu\nu} + \dots \Big]\,,
\end{align}
where the reader should notice the potential existence (depending on the specific type of string theory considered) of constant (dilaton-independent) terms involving $F^2$ terms ({\it cf.} the $c_0^{(F)}$ terms on the right-hand side of \eqref{tadpoles}).

In case we consider more general theories involving a combination of closed and open strings (the latter attached, e.g. to brane universe)s, for which one obtains effective actions in the Einstein frame that include both closed- and open-string sectors (the latter leading to DBI terms of the form appearing in the second integral on the right-hand side of \eqref{gravBI}), then, the inclusion of string loops can lead, following similar arguments to the closed-string case \eqref{effactcl}, to generalised situations, in which the (string-loop corrected) effective action acquires the form in the Einstein frame~\cite{Metsaev:1987ju}:~\footnote{Indeed, if only Abelian gauge fields are considered then only open world-sheet surfaces are taken into account, in order to evaluate the pertinent contribution to the effective action,  as discussed explicitly in \cite{Metsaev:1987ju} where it was shown that the loop corrected effective action acquires the form in the sigma-model frame (ignoring antisymmetric tensor fields contributions, which are of no interest in the present discussion): 
\begin{align}
S = \int d^4x \sqrt{-\widehat g}\, {\alpha^\prime}^{-2}e^{-2\phi} \,\Big[-\frac{3} {2}\alpha^\prime \, \big[\widehat{\mathcal R} + 4 (\nabla \phi)^2\big] + \dots   \Big]
+ d_1\, e^{-\phi} \, \sqrt{{\rm det}\Big(\widehat g_{\mu\nu} + 2\pi \alpha^\prime F_{\mu\nu}\Big)} + d_2 + d_3 + d_4\, e^\phi + \dots\,, 
\end{align}
where $d_i$, $i=1,2,\dots,$ denote finite parts of the dilaton tadpoles, and the dots denote contributions from higher derivative corrections, as well as higher string loops (that is higher powers of the string coupling $g_s=\exp(\phi)$. Passing onto the appropriate Einstein frame leads to actions of the form \eqref{loopgauge}.}
\begin{align}\label{loopgauge}
\mathcal{S} =\frac{1}{16\pi} \int \diff^4x \sqrt{-g} \Big[\mathcal{R} - 2\nabla^{\mu}\phi\nabla_{\mu}\phi \Big] -\int \diff^4x \, \sqrt{-g}\, B_{F^2}(\phi)\,  \Big[\mathcal T_4^2\, I_2^{\rm E} \Big] - 
\int \diff^4x \, \sqrt{-g}\, 
\mathcal T_4^4 \, B_{F^4}(\phi) \, I_4^{\rm E} + \dots \,,
\end{align}
where the functions $B_{F^i}(\phi), \, i=2,4$ admit a power series expansion in the string coupling, summing up terms of the generic form 
\begin{align}
 B_{F^i}(\phi) = \sum_\chi g_s^{-\chi} c_{\chi}^{(F^i)} \,,  \qquad i=1,2\,, \qquad g_s=\exp(\phi)\,,
\end{align}
where $\chi=2 - 2N - N_H$, with $N_H$ the number of holes (or boundaries) (eg disc has genus $\chi=1$, since $N=0$, $N_H=1$ etc), where 
finite parts of dilaton tadpoles contribute to the coefficients $c_\chi^{(F^i)}$.

In heterotic strings, which 
do not involve branes, the higher derivative EH electrodynamics terms do not appear in a closed BDI form. In that case a more general action, involving EH terms, after summation over string loops, might then be considered, in the Einstein frame: 
\begin{align}\label{loopgauge2}
\mathcal{S} =\frac{1}{16\pi} \int \diff^4x \sqrt{-g} \Big[\mathcal{R} - 2\nabla^{\mu}\phi\nabla_{\mu}\phi \Big] -\int \diff^4x \, \sqrt{-g}\, B_{F^2}(\phi)\,  \Big[\mathcal T_4^2\, I_2^{\rm E} \Big] - \int \diff^4 x \sqrt{-g}\, \mathcal T_4^4 B_{F^4}(\phi) \, L_{\rm EH} + \dots\,, 
\end{align}
where the ellipsis ($\dots$) includes possible string-loop generated dilaton-potential terms, whose precise form is not known at present, as this is  a highly string-model-dependent issue, and the functions
$B_{F^2}(\phi)$, $B_{F^4}(\phi)$ in this case are given by power series expansions of even powers of the string coupling, of the form \eqref{tadpoles}, as only closed world-sheet surfaces are involved. The Euler-Heisenberg Lagrangian 
$L_{\rm EH}$
is given by \eqref{EHLag}, but the coefficients $c_i, i=1,2$ 
no longer satisfy \eqref{cis}, given that the DBI action no longer describes the electromagnetic self-interactions in closed form. 
In this work we shall use the framework \eqref{loopgauge2} to discuss black hole solutions in a phenomenological manner, keeping the coefficients $c_i$ of the EH Lagrangian as arbitrary. As already mentioned, in such a framework, for some sufficiently high string-loop order one can conjecture that there would exist a {\it dilaton independent} term in the function $B_{F^4}(\phi)$. As we shall discuss in the next section, such a constant term plays a crucial r\^ole in our phenomenological analysis in this paper in yielding analytic black hole solutions of a dilaton-EH theory that could be the low energy limit of an appropriate underlying string theory. However in our work, we shall be more general, and our analysis will be presented independent of strings.

\section{Theoretical framework and black-hole solutions}\label{sec:model}

In the geometrised unit system ($c=G=1$), the Einstein-frame action functional that we will occupy us in this article is a simplified version of \eqref{loopgauge2} and reads
\begin{equation} 
\mathcal{S} =\frac{1}{16\pi} \int \diff^4x \sqrt{-g} \Big[\mathcal{R} - 2\nabla^{\mu}\phi\nabla_{\mu}\phi -e^{-2\phi}\mathcal{F}^2  -f(\phi)\big(   2\alpha\mathcal{F}^{\alpha}_{~\beta}\mathcal{F}^{\beta}_{~\gamma}\mathcal{F}^{\gamma}_{~\delta}\mathcal{F}^{\delta}_{~\alpha}-\beta \mathcal{F}^4\big)\Big]~.
\label{theory}
\end{equation}
Such a field theoretic gravitational actions also
arises as part of a non-diagonal reduction of the Gauss-Bonnet action \cite{Charmousis:2012dw} and
admits the GMGHS black hole \cite{Gibbons:1987ps, Garfinkle:1990qj} as an exact solution when $f(\phi)=0$. 
In \eqref{theory}, $\mathcal{R}$ is the Ricci scalar, $\mathcal{F}^2 \equiv \mathcal{F}_{\mu\nu}\mathcal{F}^{\mu\nu} \sim \bf{E}^{2} - \bf{B}^2$ is the usual Faraday scalar, and $\mathcal{F}^4 \equiv \mathcal{F}_{\mu\nu}\mathcal{F}^{\mu\nu}\mathcal{F}_{\alpha\beta}\mathcal{F}^{\alpha\beta}$, where $\mathcal{F}_{\mu\nu}$ stands for the usual field strength $\mathcal{F}_{\mu\nu} = \partial_{\mu}\mathcal{A}_{\nu} - \partial_{\nu}\mathcal{A}_{\mu}$ and $\alpha,\beta$ are coupling constants of the theory, with dimensions (length)$^2$, which in our discussion are treated phenomenologically.
The scalar field $\phi$ and the associated scalar function $f(\phi)$ are both dimensionless.\footnote{The reader should be reminded at this stage that in the special case of (open)string/brane-inspired BI theory at tree-level in string loops, the function $f(\phi) \sim e^{-5\phi}$ ({\it cf.} \eqref{gravBI}), however in such a case the Maxwell term $\mathcal F^2$ in \eqref{theory} should be accompanied by the inverse of the open string coupling, {\it ie.} $e^{-\phi}$, instead of $e^{-2\phi}$ that appears in \eqref{theory}. On the other hand, in the heterotic-string-inspired model \eqref{theory} can be mapped to the model \eqref{theory2}, upon choosing $\gamma=1$, and $\mathcal F \widetilde{\mathcal F}=0$, that is concentrating on magnetically charged black holes only (in which case, the function $f(\phi) \sim e^{-6\phi}$). However, as we have already stressed, and we shall argue below, it is crucial for an analytic treatment of the black-hole solution to have a dilaton-independent term in $f(\phi)$, which, as we have argued in the previous section, can be induced by considering higher-order string loop corrections in the underlying string-theory model.}    
For the moment we do not consider a potential for the dilaton, but only its non-linear interactions with the EH terms. The addition of a pure dilaton potential $\mathfrak{V}(\phi)$ can lead to interesting alternative solutions, including a cosmological constant, 
which we discuss in section \ref{sec:othersol}. 

The field equations emanating from \eqref{theory} are of the following form
\begin{align}
\label{grav-eqs}
 G_{\mu\nu} =&\ \nonumber 2\partial_{\mu}\phi\partial_{\nu}\phi - g_{\mu\nu}\partial^{\alpha}\phi\partial_{\alpha}\phi + 2 e^{-2\phi} \left( \mathcal{F}_{\mu}^{~ \alpha}\mathcal{F}_{\nu\alpha} - \frac{1}{4}g_{\mu\nu}\mathcal{F}^2\right) +\\&f(\phi)\left\{8  \alpha \mathcal{F}_{\mu}^{~\alpha}\mathcal{F}_{\nu}^{~\beta}\mathcal{F}_{\alpha}^{~\eta}\mathcal{F}_{\beta\eta} - \alpha  g_{\mu\nu}\mathcal{F}^{\alpha}_{~\beta}\mathcal{F}^{\beta}_{~\gamma}\mathcal{F}^{\gamma}_{~\delta}\mathcal{F}^{\delta}_{~\alpha}  
- 4 \beta  \mathcal{F}_{\mu}^{~\xi}\mathcal{F}_{\nu\xi}\mathcal{F}^2 + \frac{1}{2}g_{\mu\nu}\beta \mathcal{F}^4\right\}~,
\end{align}
\begin{equation}
    4\square \phi = -2e^{-2\phi}\mathcal{F}^2 +\frac{df(\phi)}{d\phi}\left(2\alpha\mathcal{F}^{\alpha}_{~\beta}\mathcal{F}^{\beta}_{~\gamma}\mathcal{F}^{\gamma}_{~\delta}\mathcal{F}^{\delta}_{~\alpha}-\beta \mathcal{F}^4\right)~,
\end{equation}
\begin{equation}
    \partial_\mu\Big\{ \sqrt{-g} \Big[4 \mathcal{F}^{\mu\nu}\left(2\beta f(\phi)\mathcal{F}^2-e^{-2\phi}\right)
    -16 \alpha \mathcal{F}^{\mu}{}_\kappa \mathcal{F}^{\kappa}{}_\lambda \mathcal{F}^{\nu\lambda}\Big] \Big\}=0\,.
\end{equation}
By taking into account the higher-order electromagnetic invariants $\mathcal{F}^4$ and $\mathcal{F}^{\alpha}_{~\beta}\mathcal{F}^{\beta}_{~\gamma}\mathcal{F}^{\gamma}_{~\delta}\mathcal{F}^{\delta}_{~\alpha}$, we are interested in extending the GMGHS solution \cite{Gibbons:1987ps, Garfinkle:1990qj}.
To do so, we introduce the most general spherically symmetric metric ansatz in the form
\begin{equation}
\label{line-elem}
\diff s^2 = -B(r)\diff t^2 + \frac{\diff r^2}{B(r)} + [R(r)]^2 \diff\Omega^2~, 
\end{equation}
where $B(r),R(r)$ are two unknown functions to be determined from the field equations, while $\diff\Omega^2=\diff\theta^2+\sin^2\theta \diff\varphi^2$.\,\footnote{Note that throughout this article, $\varphi$ will always denote the azimuthal coordinate, while $\phi$ will always denote the scalar field.} 
Moreover, we consider both electric and magnetic charges, via the following four-vector, which is compatible with spherical symmetry,
\begin{equation} \mathcal{A}_{\mu} = (V(r),0,0,Q_m \cos \theta)~,\end{equation}
where $Q_m$ stands for the magnetic charge carried by the black hole. 
This ansatz for the electromagnetic field solves by construction the $\varphi$ component of the Maxwell equations iff one considers that the scalar field inherits the spacetime symmetries, namely $\phi \equiv \phi(r)$. 
Interestingly, one can see that the combination
\begin{equation}2\alpha\mathcal{F}^{\alpha}_{~\beta}\mathcal{F}^{\beta}_{~\gamma}\mathcal{F}^{\gamma}_{~\delta}\mathcal{F}^{\delta}_{~\alpha} -\beta\mathcal{F}^4 =\frac{4 (\alpha -\beta ) Q_m^4}{[R(r)]^8}+\frac{8 \beta  Q_m^2 [V'(r)]^2}{[R(r)]^4}+4 (\alpha -\beta ) [V'(r)]^4~,\end{equation}
will vanish if one does not consider both electric and magnetic configurations in the case of $\alpha=\beta$.
In the above, prime denotes derivation with respect to $r$.
Maxwell's equation is very difficult to be integrated for the dyonic case and as a result we will consider pure magnetic fields, that is $V(r)=0$. 
Consequently, both these non-linear electrodynamics terms will contribute iff $\alpha\neq\beta$.
We will begin our analysis for the scalar free scenario $\phi=0, f(\phi=0)=1$, for which the solution reads
\begin{equation}
\label{phi-free-sol}
B(r)=1-\frac{2 M}{r} +\frac{Q_m^2}{r^2}+\frac{2  (\alpha-\beta)  Q_m^4}{5 r^6}~,
\end{equation}
and $R(r)=r$. This solution resembles the Einstein-Euler-Heisenberg black hole \cite{Yajima:2000kw}. The interesting thing to notice in \eqref{phi-free-sol} is that the non-linear electromagnetic terms $\mathcal{F}^{\alpha}_{~\beta}\mathcal{F}^{\beta}_{~\gamma}\mathcal{F}^{\gamma}_{~\delta}\mathcal{F}^{\delta}_{~\alpha}$ and $\mathcal{F}^4$ affect the spacetime geometry in a similar way.
It is solely the values of the coupling constants $\alpha$ and $\beta$ that determine whether this contribution survives or not.
Note that in the case of $\alpha=\beta$ the higher-order electromagnetic term does not contribute at all.
However, in the case where $\alpha \neq \beta$, we notice that depending on the signs of the parameters $\alpha$ and $\beta$, the non-linear electromagnetic terms can act either attractively or repulsively.
Black holes with a scalar hair in the Euler-Heisenberg theory have been discussed in \cite{Karakasis:2022xzm}, and it was found that the scalar hair results in a more compact black hole (having a smaller radius for the event horizon) when compared to the non-hairy Einstein-Euler-Heisenberg black hole. 

Let us now assume a non-trivial profile for the coupling function $f(\phi)$. In particular, we consider
\begin{equation} f(\phi) = -[3 \cosh (2 \phi )+2] \equiv - \frac{1}{2} \left(3 e^{-2 \phi } + 3 e^{2 \phi }+4\right)~. \label{couplingfunction}
\end{equation}
Notice here that the coupling function $f(\phi)$ contains the dilatonic coupling $e^{2\xi\phi}$ with $\xi=\pm 1$ as well as a constant (dilaton independent) term. At this point the reader is invited, for completion, to compare such couplings with the string-loop corrected coupling functions $B_{F^4}(\phi)$, in the framework of string-inspired models \eqref{loopgauge2}, discussed in the previous section \ref{sec:ned}. In such a stringy context, the exponential dilaton terms in the coupling function \eqref{couplingfunction} can be written as $f(\phi)= -\frac{3}{2} (g_s^{-2} + g_s^2) -2$, where $g_s=\exp(\phi)$ is the string coupling. As discussed in section \ref{sec:ned},  the $g_s^{-2}$ is the standard tree-level dilaton-Maxwell term coupling~\cite{Gross:1986mw,Green:2012oqa,Green:2012pqa}, while the $g_s^2$ indicates two-string-loop corrections (genus-$\chi=2$ world-sheet surfaces). The crucial, for our subsequent discussion, dilaton-independent term in $f(\phi)$ might be the result of appropriate combinations of higher-string-loop corrections in the Einstein-frame effective action. 

It is now straightforward to solve the field equations of \eqref{theory}, with \eqref{couplingfunction}, in order to determine the geometry of the spacetime and the functional expression for the scalar field.
By doing so, one obtains a simple exact, magnetically charged black-hole solution, for which it holds that
\begin{gather}
\label{B,R-r}
B(r) =1-\frac{2 M}{r}-\frac{2 (\alpha -\beta ) Q_m^4}{r^3 \left(r-\frac{Q_m^2}{M}\right)^3}~,\hspace{1.5em}[R(r)]^2 =r \left(r-\frac{Q_m^2}{M}\right)~,\\[2mm]
\phi(r) = -\frac{1}{2}\ln \left(1-\frac{Q_m^2}{Mr}\right)~,\hspace{1.5em}\mathcal{A}_\mu=(0,0,0,Q_m \cos\theta)~.
\label{fmn}
\end{gather}
We observe that in this case, for $\alpha=\beta$ we obtain the GHS solution \cite{Garfinkle:1990qj}, while the radial coordinate $r\in(Q_m^2/M,+\infty)$ in order to have $R\in(0,+\infty)$. 
In this case, it is also intriguing to observe that the sign of the combination $\alpha-\beta$ among the coupling constants determines whether the higher-order electromagnetic terms in the theory will contribute attractively or not.

To obtain a better understanding of the spacetime geometry, one may express the line element \eqref{line-elem} in terms of the physical coordinate system with $R$ playing the role of the radial coordinate.
By doing so, one finds that
\begin{gather}
    \label{line-elem-phys}
    ds^2=-B(R)dt^2+\frac{[W(R)]^2dR^2}{B(R)}+R^2 d\Omega^2\,,
\end{gather}
with functions $B(R)$, $W(R)$, and $\phi(R)$ being given by
\begin{gather}
    \label{B-phys}
    B(R)=1-\frac{4M^2}{Q_m^2 + \sqrt{Q^4_m+4M^2R^2}}-\frac{2(\alpha-\beta)Q_m^4}{R^6}\,,\\[2mm]
    \label{W-phys}
    [W(R)]^2= \frac{4M^2R^2}{Q^4_m+4M^2R^2}\,,\\[2mm]
    \label{phiphys}
    \phi(R) = -\frac{1}{2} \ln \left(\frac{\sqrt{Q_m^4+4 M^2 R^2}-Q_m^2}{\sqrt{Q_m^4+4 M^2 R^2}+Q_m^2}\right)\,.
\end{gather}
In the physical coordinate system $(t,R,\theta,\varphi)$, one can verify that the curvature invariant quantities $\mathcal{R}$, $\mathcal{R}_{\mu\nu}\mathcal{R}^{\mu\nu}$, and $\mathcal{R}_{\alpha\beta\gamma\delta}\mathcal{R}^{\alpha\beta\gamma\delta}$ possess a single spacetime singularity residing at $R=0$, while the function $B(R)$ satisfies the following expansions
\begin{gather}
    \label{B-R-exp-inf}
    B(R\rightarrow +\infty)=1-\frac{2M}{R}+\frac{Q_m^2}{R^2}-\frac{Q_m^4}{4M R^3}+\frac{Q_m^8}{64M^3R^5}-\frac{2(\alpha-\beta)Q^4_m}{R^6}+\mathcal{O}(1/R^7)\,,\\[2mm]
    \label{B-R-exp-0}
    B(R\rightarrow 0)=-\frac{2(\alpha-\beta)Q_m^4}{R^6}+\left(1-\frac{2M^2}{Q_m^2}\right)+\mathcal{O}(R^2)\,.
\end{gather}
From \eqref{B-R-exp-inf}, it becomes apparent that the spacetime \eqref{line-elem-phys} is practically indistinguishable from that of a magnetically charged Reissner-Nordstr\"{o}m black hole for an observer at infinity, with the parameter $M$ corresponding to the ADM mass of the solution.
However, an observer much closer to the black hole \eqref{line-elem-phys} would perceive a completely different picture.
Indeed these quantum-gravity corrections are important near the singularity, since the geometry there is determined by their behavior.

%%%%%%%%%%%%%%%%%%%%%%
\begin{figure}[t]
    \centering
    \begin{subfigure}[b]{0.49\textwidth}
    \includegraphics[width=1\textwidth]{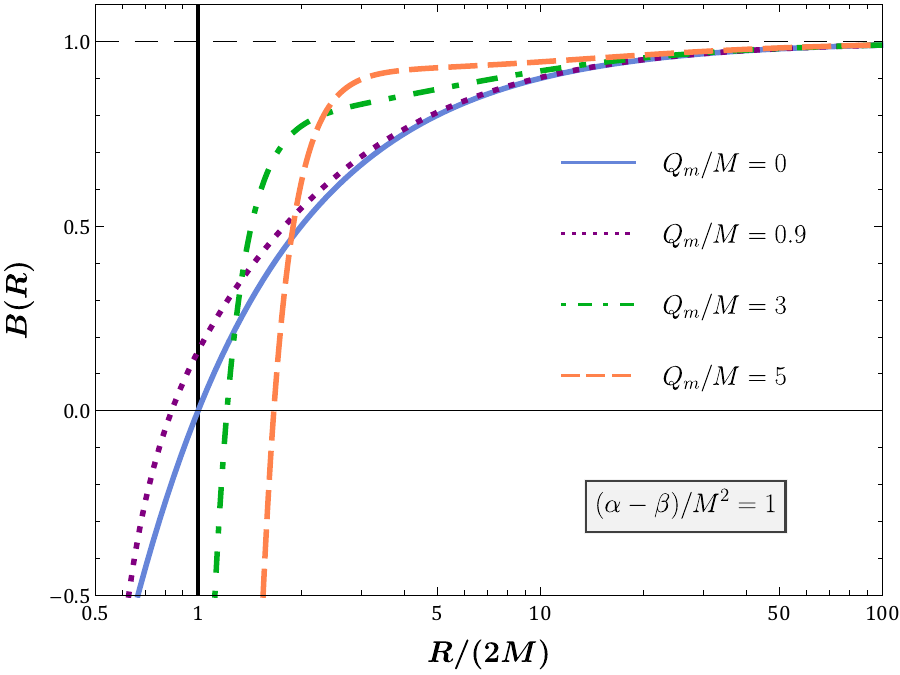}
    \caption{\hspace*{-2.5em}}
    \label{subf: bh-hor1}
    \end{subfigure}
    \hfill
    \begin{subfigure}[b]{0.48\textwidth}
    \includegraphics[width=1\textwidth]{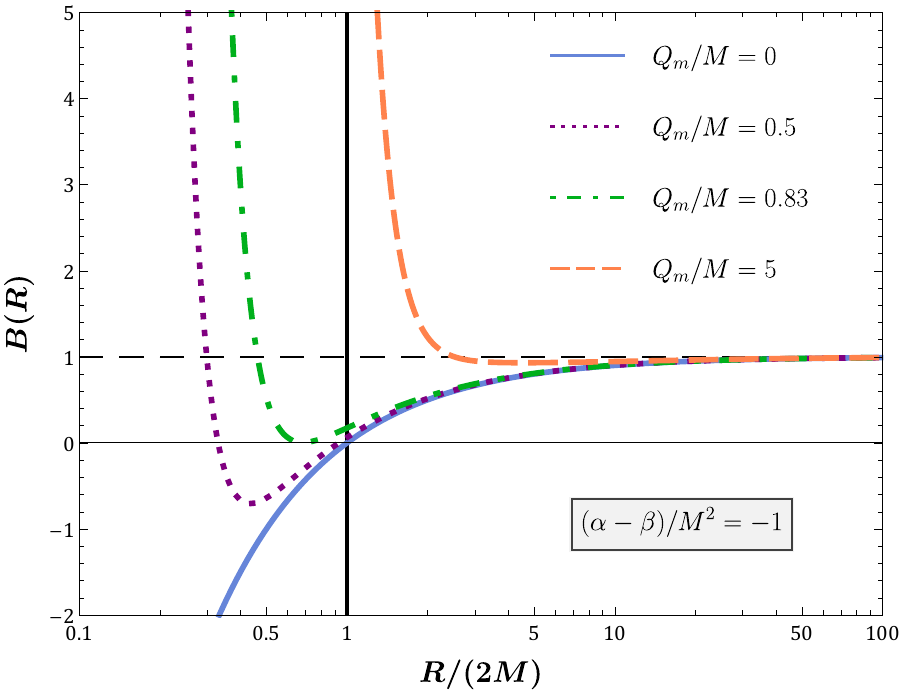}
    \caption{\hspace*{-2.2em}}
    \label{subf: bh-hor2}
    \end{subfigure}
    \caption{The metric function $B(R)$ in terms of $R/(2M)$ for various values of the parameter $Q_m/M$ and (a) $(\alpha-\beta)/M^2=1$, (b) $(\alpha-\beta)/M^2=-1$. All parameters are dimensionless, and the horizontal axis in both figures is logarithmic.}
    \label{fig: bh-hor}
\end{figure}
%%%%%%%%%%%%%%%%%%%%%%

The radial null-trajectories for the spacetime \eqref{line-elem-phys}, lead to the relation
\begin{equation}
    \frac{dt}{dR}=\pm \frac{2MR}{\sqrt{Q^4_m+4M^2R^2}}\frac{1}{|B(R)|}\,,
\end{equation}
which by its turn means that the roots of the function $B(R_h)$ correspond to black-hole horizons.
In Fig. \ref{fig: bh-hor}, one can observe the behavior of the metric function $B(R)$ in terms of the dimensionless quantity $R/(2M)$.
We see that the solution \eqref{line-elem-phys} describes a black hole with a single horizon when $(\alpha-\beta)/M^2=1$, while for $(\alpha-\beta)/M^2=-1$ the black-hole horizons can range from two to none.
It is essential to note that the previous assertion holds in general for $(\alpha-\beta)/M^2$ being either greater or lower than zero.
Analysis of Fig. \ref{subf: bh-hor1} reveals that a positive value for the combination $(\alpha-\beta)/M^2$ results in black-hole solutions featuring a single horizon. 
To facilitate comparison, we have also included the Schwarzschild solution which can be obtained by simply setting $Q_m=0$. 
One can readily observe that within our theory's solution spectrum, black holes can exhibit either greater compactness or sparsity relative to the Schwarzschild solution.
In astrophysical scenarios where $Q_m$ is relatively small compared to the mass, our solution appears more compact.
Conversely, when the fraction $(\alpha-\beta)/M^2$ takes a negative value, the solutions range from black holes with two horizons to naked singularities. 
The transition from one class of solutions to the other occurs continuously as the magnetic charge $Q_m$ increases, as depicted in Fig. \ref{subf: bh-hor2}. 
Consequently, in this scenario, there always exists a specific value for the ratio $Q_m/M$ that renders the black hole extremal, meaning the inner and outer horizons coincide.

%%%%%%%%%%%%%%%%%%%%%%%%%%%%%%%%%
\begin{figure}[t]
    \centering
    \includegraphics[width=0.5\textwidth]{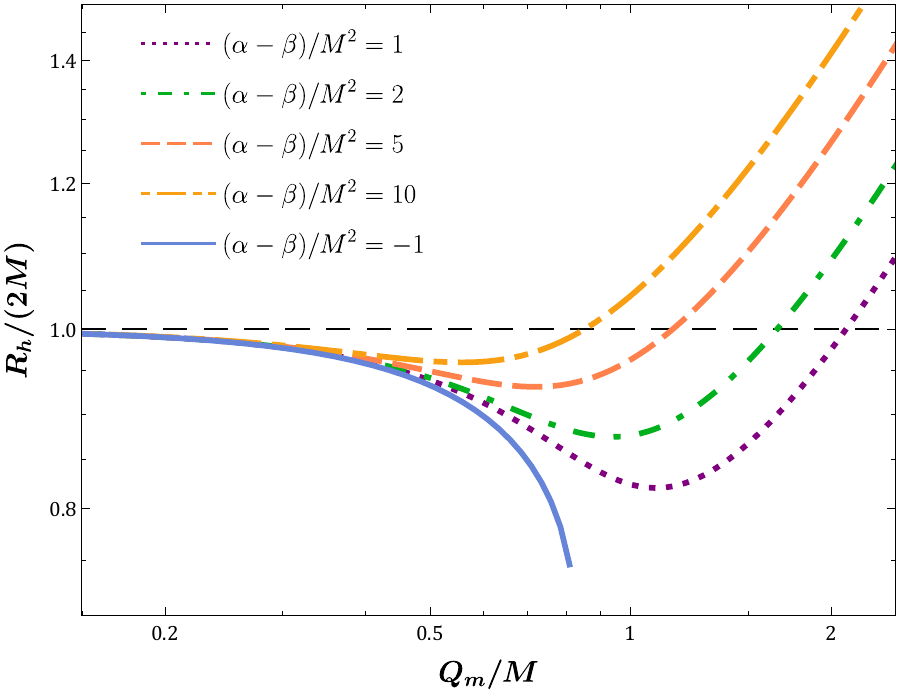}
    \caption{The ratio $R_h/(2M)$ in terms of the ratio $Q_m/M$ for various values of the dimensionless parameter $(\alpha-\beta)/M^2$. Both axes are logarithmic. }
    \label{fig: Rh-Qm-M}
\end{figure}
%%%%%%%%%%%%%%%%%%%%%%%%%%%%%%%%%

It is crucial to highlight here the intriguing behavior observed in the realm of single-horizon black-hole solutions, for which $\alpha-\beta>0$. 
Specifically, there exists a minimum value for the ratio $R_h/(2M)$, which is below unity, resulting in more compact black holes compared to the Schwarzschild solution.
Starting from $Q_m=0$ (Schwarzschild) and increasing the magnetic charge, the resulting black holes become progressively more compact until reaching the point where $R_h/(2M)$ attains its minimum value.
Beyond this point, further increase in the ratio $Q_m/M$ causes $R_h/(2M)$ to rise again, eventually reaching $R_h/(2M)=1$, albeit now with $Q_m\neq 0$.
Subsequently, any additional increase in the ratio $Q_m/M$ yields a solution more sparse than the Schwarzschild counterpart.
This particular behavior is elucidated by analyzing Fig. \ref{fig: Rh-Qm-M}, where the relationship between the ratio of the black hole horizon ($R_h$) to twice the black hole mass ($2M$) and the ratio $Q_m/M$ is depicted for various values of the dimensionless parameters $(\alpha-\beta)/M^2$.
Conversely, it is observed that when $\alpha-\beta<0$, the outer horizon radius of the resulting black holes is consistently smaller than that of the corresponding Schwarzschild black hole with the same mass. 
Furthermore, it is important to note that in this scenario, the graph reaches a termination point. This occurs because, beyond a certain threshold of the ratio $Q_m/M$ (which is less than unity), there is a significant transition in the nature of the compact object. 
Specifically, the object transitions from being an extremal black hole to a naked singularity. 
Consequently, for this particular choice of parameters, there is no horizon to be depicted. These observations are further corroborated by the findings depicted in Fig. \ref{subf: bh-hor2}.
Returning now to the case $\alpha-\beta>0$, the discovery of black-hole solutions sharing identical horizon radii yet varying in the ratios $Q_m/M$ unveils a realm of \emph{doppelg\"{a}nger black holes} within the framework of theory \eqref{theory}.
While it is typical to find black holes stemming from different theoretical paradigms with shared horizon radii but differing physical attributes such as mass, electric charge, or secondary scalar hair, such occurrences are notably rare when considering black holes that arise from the same theory.
Even more remarkable is the fact that these two doppelg\"{a}nger black holes, despite having identical horizon radii, exhibit distinguishable thermodynamic behaviors. One is thermodynamically stable while the other is unstable. This distinctive feature is thoroughly explored in Section \ref{sec:thermo}.

\section{Geodesics and energy conditions}
\label{sec:geodener}

\subsection{Geodesics}

In this subsection, we will examine the geodesic curves of massive particles and the effective gravitational potential generated by the spacetime geometry given by eqs. \eqref{line-elem} and \eqref{B,R-r}.
We choose to work with the $(t,r,\theta,\varphi)$ coordinate system, as it facilitates a straightforward derivation of the effective gravitational potential $V_{\text{eff}}$ through a well-established procedure.
This will help us to better comprehend the geometry of the aforementioned black hole solutions.
To do so, we introduce the effective Lagrangian
\begin{gather}
    \label{geo}
    2L_{\text{eff}} =g_{\mu\nu}\frac{dx^{\mu}}{d\tau}\frac{dx^{\nu}}{d\tau}=-B(r)\dot{t}^2+\frac{\dot{r}^2}{B(r)}+[R(r)]^2\left(\dot{\theta}^2+\sin^2\theta\, \dot{\varphi}^2\right)\,,   
\end{gather}
the Euler-Lagrange equations of which yield the geodesic equations.
In the above, $\tau$ is an affine parameter of motion which can be identified with the proper time of a particle, dot denotes derivation with respect to $\tau$, while $2L_{\text{eff}}=-1$ corresponds to massive particles which follow a timelike path. Note that massless particles will not follow the geodesics induced by the geometry $g_{\mu\nu}$, instead they will follow the geodesics induced by an effective geometry that accounts for photon-photon interactions, introduced by the non-linear electromagnetic terms in our action. 
Upon inspecting the Lagrangian (\ref{geo}), it becomes evident that there is no explicit dependence on the coordinates $(t, \varphi)$. 
As a result, the Euler-Lagrange equations for $t$ and $\varphi$ yield two conserved quantities: the energy $E$ and the angular momentum $J$ of the particle under consideration, respectively.
Hence, we have
\begin{eqnarray}
    && E = B(r) \dot{t}~,\\
    && J =  [R(r)]^2\sin^2\theta\, \dot{\varphi}~.
\end{eqnarray}
The equation of motion for $\theta$ reads
\begin{equation}
    [R(r)]^2\,\ddot{\theta} + 2R(r)R'(r)\,\dot{r}\dot{\theta} - J\, \frac{\cos\theta }{\sin\theta}=0~,
\end{equation}
and by choosing $\theta=\pi/2$ ($\dot{\theta}=0$), the particles stay fixed at the equatorial plane. Now plugging these results back to (\ref{geo}) we obtain the radial equation of motion
\begin{equation}
    \frac{1}{2}\dot{r}^2+V_{\text{eff}}(r)=\frac{1}{2}E^2,
\end{equation}
with the effective potential induced by the geometry being
\begin{equation}
    V_{\text{eff}}(r) = \frac{B(r)}{2} \left(1+\frac{J^2}{[R(r)]^2}\right)~, \label{V1}
\end{equation}
and the functions $B(r)$ and $R(r)$ given by \eqref{B,R-r}\,.
As we have already mentioned in the previous section, the radial coordinate $r$ ranges from $Q_m^2/M$ to plus infinity because the physical radial coordinate $R\in(0,+\infty)$.

%%%%%%%%%%%%%%%%%%%%%%%%%%%%%%%%%
\begin{figure}[t]
    \centering
    \includegraphics[width=0.5\textwidth]{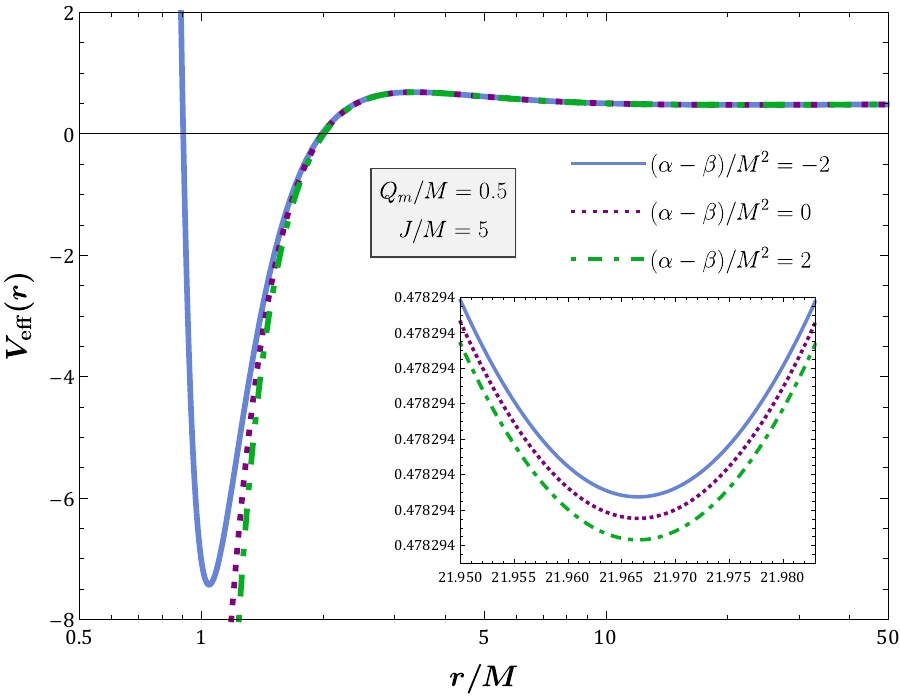}
    \caption{The effective potential $V_{\text{eff}}$ in terms of the quantity $r/M$ for $Q_m/M=0.5$, $J/M=5$, and $(\alpha-\beta)/M^2=\{-2,0,2\}$. All parameters are dimensionless, and the horizontal axis is logarithmic.}
    \label{fig: Veff}
\end{figure}
%%%%%%%%%%%%%%%%%%%%%%%%%%%%%%%%%

In Fig. \ref{fig: Veff}, we depict the behavior of the effective potential $V_{\text{eff}}$ in terms of the dimensionless parameter $r/M$, considering three distinct values for the fraction $(\alpha-\beta)/M^2$.
Upon close examination, it becomes evident that the scenarios where $\alpha=\beta$ and $(\alpha-\beta)/M^2=2$ share a strikingly similar pattern in the effective potential. In both cases, the potential curve features one maximum and one minimum value, corresponding to unstable and stable circular orbits, respectively.
On the other hand, in the case where $(\alpha-\beta)/M^2=-2$, an additional minimum emerges, exhibiting local behavior that closely resembles the Newtonian potential.
To understand the origin of this difference, we have to examine the expansion of the potential $V_{\text{eff}}$ in the limit $r\rightarrow Q_m^2/M$, where one can verify that
\begin{equation}
    V_{\text{eff}}(r\rightarrow Q_m^2/M)=-\frac{J^2M^4 (\alpha -\beta )}{Q_m^4 \left(r-\frac{Q_m^2}{M}\right)^4}+\frac{M^3 (\alpha -\beta ) \left(4 J^2 M^2-Q_m^4\right)}{Q_m^6 \left(r-\frac{Q_m^2}{M}\right)^3}+\mathcal{O}\Bigg[\left(r-\frac{Q_m^2}{M}\right)^{-2} \Bigg]\,.
\end{equation}
We observe that the first term, which dominates in this particular regime, depends explicitly on the sign of the quantity $\alpha-\beta$. 
When $\alpha-\beta>0$, the potential tends toward negative infinity, whereas for $\alpha-\beta<0$, the potential tends toward positive infinity. This alignment precisely mirrors our observations in Fig. \ref{fig: Veff}.
Finally, from Fig. \ref{fig: Veff}, it is also clear that for $r/M>2$, the effective potential in all cases exhibits the same profile, independently of the relative values of the coupling constants $\alpha$ and $\beta$.
This can be naively understood through the expansion of the potential at infinity, which is of the following form 
\begin{equation}
    V_{\text{eff}}(r\rightarrow +\infty)= \frac{1}{2}-\frac{M}{r}+\frac{J^2}{2 r^2}+\left(\frac{Q_m^2}{2M^2}-1\right)\left[\frac{J^2M}{r^3}+\frac{J^2Q_m^2}{r^4}+\frac{J^2Q_m^4}{M\,r^4}+\frac{J^2Q_m^6}{M^2r^4}\right]-\frac{(\alpha-\beta)  Q_m^4}{r^6}+\mathcal{O}\left(\frac{1}{r^7}\right)\,.
\end{equation}
It is obvious that in the asymptotic regime, the coupling constants $\alpha$ and $\beta$ cease to influence the potential profile, as their first contribution comes into play only in the seventh term of the expansion. 
Consequently, even at medium distances, we anticipate that beyond a certain point, the coupling constants will have negligible impact on the potential's behavior.

\subsection{Energy Conditions}
We will now turn our attention to the energy conditions associated with the stress-energy tensor of our theory.
In the physical coordinate system $(t,R,\theta,\varphi)$ the stress-energy tensor is described by an anisotropic fluid which in a covariant form can be written as
\begin{equation}
    \label{en-mom-fl}
    T^{\mu\nu}=(\rho_E+p_\theta)u^\mu u^\nu+(p_R-p_\theta)n^\mu n^\nu+p_\theta g^{\mu\nu}\,.
\end{equation}
%

%%%%%%%%%%%%%%%%%%%%%%
\begin{figure}[t]
    \centering
    \begin{subfigure}[b]{0.48\textwidth}
    \includegraphics[width=1\textwidth]{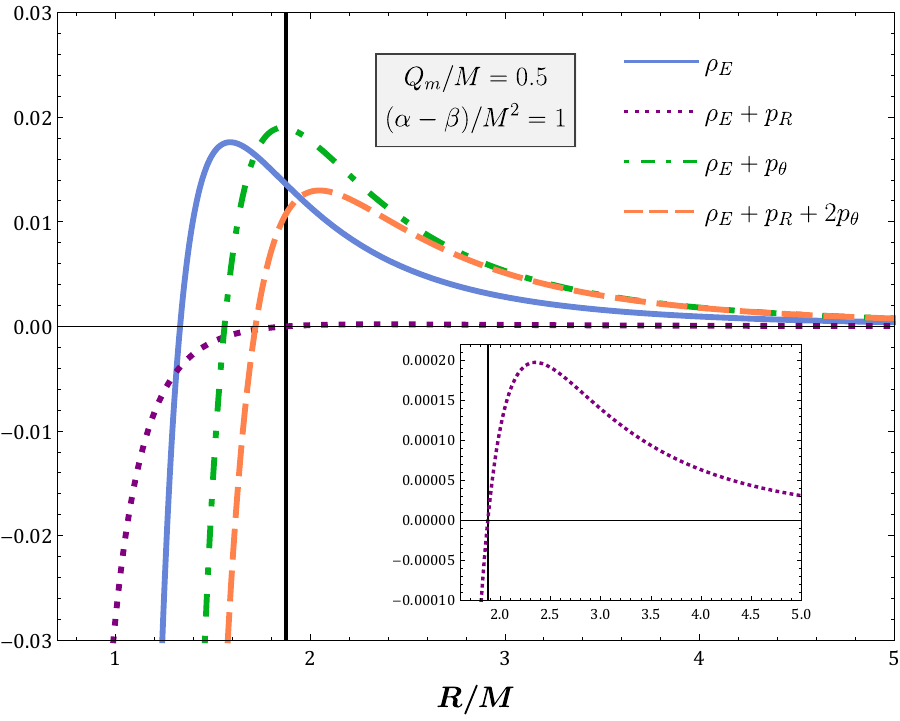}
    \caption{\hspace*{-2em}}
    \label{subf: encon1}
    \end{subfigure}
    \hfill
    \begin{subfigure}[b]{0.48\textwidth}
    \includegraphics[width=1\textwidth]{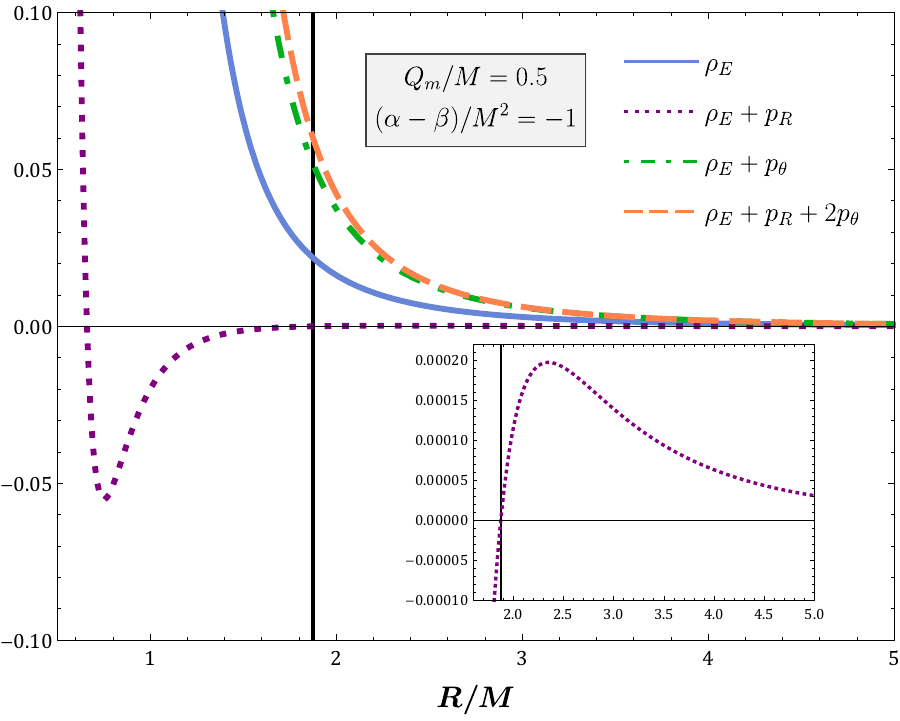}
    \caption{\hspace*{-1.8em}}
    \label{subf: encon2}
    \end{subfigure}
    \caption{The energy conditions for $Q_m/M=0.5$ with (a) a positive and (b) a negative value assigned to the dimensionless quantity $(\alpha-\beta)/M^2$. The vertical lines correspond to the horizon of the black hole determined by these parameters.}
    \label{fig: EC}
\end{figure}
%%%%%%%%%%%%%%%%%%%%%%

In the above, $\rho_E$ is the energy density of the fluid measured by a comoving observer with the fluid, $p_R$ is its radial pressure, $p_\theta$ is its tangential pressure, while $u^\mu$ and $n^\mu$ 
are its timelike four-velocity and a spacelike unit vector orthogonal to $u^\mu$ and also to both angular directions.
The four-vectors $u^\mu$ and $n^\mu$ satisfy the following relations:
\begin{align}
    \label{u-4v}
    &u^\mu=u(R)\, \delta^{\mu}_0\,, \hspace{1em}u^\mu u^\nu g_{\mu\nu}=-1\,,\\[2mm]
    \label{n-4v}
    &n^\mu=n(R)\, \delta^{\mu}_1\,, \hspace{1em}n^\mu n^\nu g_{\mu\nu}=1\,.
\end{align}
Given eqs. \eqref{grav-eqs}, \eqref{couplingfunction}, \eqref{line-elem-phys}-\eqref{phiphys}, and \eqref{en-mom-fl}-\eqref{n-4v}, one can readily compute that
\begin{align}
    \rho_E=-T^{t}{}_t=\frac{B(R)}{[W(R)]^2}\left(\frac{\diff \phi}{\diff R}\right)^2+\frac{Q_m^2}{R^4}\, e^{-2\phi}+\frac{2(\alpha-\beta)Q_m^4}{R^8}\, f(\phi)\,,\label{Ttt}
    \end{align}
    \begin{align}
    p_R= T^{R}{}_R=\frac{B(R)}{[W(R)]^2}\left(\frac{\diff \phi}{\diff R}\right)^2-\frac{Q_m^2}{R^4}\, e^{-2\phi}-\frac{2(\alpha-\beta)Q_m^4}{R^8}\, f(\phi)\,,\label{TRR}
    \end{align}
    \begin{align}
    p_\theta = T^{\theta}{}_\theta=T^{\varphi}{}_\varphi=-\frac{B(R)}{[W(R)]^2}\left(\frac{\diff \phi}{\diff R}\right)^2+\frac{Q_m^2}{R^4}\, e^{-2\phi}+\frac{6(\alpha-\beta)Q_m^4}{R^8}\, f(\phi)\,.
    \label{Tthetatheta}
\end{align}

For the anisotropic fluid of \eqref{en-mom-fl}, the energy conditions take the following expression:
\begin{itemize}
\item{Null Energy Conditions (NEC): 
$\hspace{1.7em}\rho_E+p_R \geq 0 \hspace{1em} \&\hspace{1em} \rho_E+p_\theta \geq 0\,,$}
\item{Weak Energy Conditions (WEC): 
$\hspace{1em}\text{NEC} \hspace{1em}\&\hspace{1em} \rho_E\geq 0\,,$}
\item{Strong Energy Conditions (SEC): 
$\hspace{1em}\text{NEC} \hspace{1em}\&\hspace{1em} \rho_E+p_R+2p_\theta \geq 0\,.$
}
\end{itemize}
In Figs. \ref{subf: encon1} and \ref{subf: encon2}, we illustrate the graphs of the quantities $\rho_E$, $\rho_E+p_R$, $\rho_E+p_\theta$, and $\rho_E +p_R+2p_\theta$ each plotted against the dimensionless parameter $R/M$. 
The free parameters of our model and solution have been chosen to be $Q_m/M=0.5$, while the combination $(\alpha-\beta)/M^2$ takes values of $1$ and $-1$, respectively.
It is evident from Fig. \ref{fig: EC} that all the aforementioned quantities maintain positive values within the causal region of spacetime and as a result, all energy conditions are satisfied. 

These results imply, therefore, the existence of a dilaton hair in the black hole's exterior, while the energy conditions are satisfied, 
thereby leading to a bypass of the pertinent (modern version of the) no-hair theorems~\cite{Bekenstein:1995un,Sotiriou:2013qea} in the spirit of \cite{Dorlis:2023qug}. 
The situation can be understood as a consequence of the fact that the stress-energy tensor of our theory \eqref{theory}, with \eqref{couplingfunction},
is such that the tangential component of the pressure ($p_\theta = T^\theta_{\,\,\theta}$) dominates over its radial one ($p_R = T^R_{\,\,R})$ (in the $(t,R,\theta,\phi)$ coordinate system), outside the horizon. That is, the following quantity is positive 
in the exterior region of the black hole,
\begin{align}\label{gtt}
\mathcal G - \mathcal J = T^\theta_{\,\,\theta} - T^R_{\,\,R} > 0\,,
\end{align}
where $\mathcal G = \rho_E + T^\theta_{\,\,\theta}$ and $\mathcal J \equiv \rho_E + T^R_{\,\,R}$. Note that the condition \eqref{gtt} follows from NEC.
As discussed in detail in \cite{Dorlis:2023qug}, the quantity $2\mathcal G/R$ is the {\it effective gradient pressure force}, and its positivity (i.e. that of $\mathcal G$, since $R >0$) explains in a physical way the existence of 
scalar hair in the black-hole's exterior, without any violation of 
the energy conditions. The validity of the condition \eqref{gtt} can also be explicitly checked in our model from Eqs.~\eqref{TRR} and \eqref{Tthetatheta}. Thus, the exact black hole solution of the self-gravitating scalar-EH (non-linear) electrodynamics examined in this paper constitutes another explicit example of the general considerations of \cite{Dorlis:2023qug} for bypassing the no-hair theorem without any violation of the energy conditions.

\section{Thermodynamic analysis}\label{sec:thermo}

In this section, we will discuss the thermodynamics of both the GMGHS and our black-hole solution by considering their Euclidean actions. 
We will consider the Grand Canonical Ensemble and enclose the black hole spacetime in a cavity with a large radius $r_c$. 
In the Grand Canonical Ensemble, the black hole is allowed to exchange energy/mass and charge with its environment, so these two quantities are allowed to flow in and out through the boundary keeping the temperature and the magnetostatic potential of the boundary fixed. 
This effectively means that $T(r_h) = T(r_c)$ and $\Phi_m(r_h) = \Phi_m(r_c)$ and the system black hole-cavity is in thermodynamic equilibrium. 
Note that $T$ is the black-hole temperature and $\Phi_m$ is the magnetostatic potential.
The quantum partition function for the system is then    given by 
\begin{equation}
    \label{part-fun}
    \mathcal{Z} = \int \diff[g^{(L)}_{\mu\nu},\psi]e^{i\mathcal{S}(g^{(L)}_{\mu\nu},\psi)} =\int \diff[g^{(E)}_{\mu\nu},\psi]e^{-\mathcal{I}_{E}(g^{(E)}_{\mu\nu},\psi)}~,
\end{equation}
where $\mathcal{S}$ is the Lorentzian action, $\mathcal{I}_E$ is the Euclidean action and $\psi$ denotes all other possible fields included besides the metric tensor. 
The two actions are related via $\mathcal{I}_E = -i \mathcal{S}$ \cite{Bravo-Gaete:2014haa}. 
The quantity $g_{\mu\nu}^{(L)}$ is the Lorentzian metric with signature $(-+++)$, which corresponds to a $\mathbb{R}^{3,1}$ spacetime, while $g_{\mu\nu}^{(E)}$ is the Euclidean metric with signature $(++++)$, which is obtained from the Lorentzian one by performing a Wick rotation \cite{FULLING1987135} of the time coordinate ($\tau=i t$). In the 
standard Matsubara formalism of finite-temperature systems, the Euclidean metric 
corresponds to a space $\mathbb{R}^3\times S_{\beta_{\tau}}^1$, where the radius $\beta_{\tau}$ of the $S^1$ is the inverse temperature $T^{-1}$ in units of the Boltzman factor $k_B=1$.
Hence, the second integral in \eqref{part-fun} is evaluated over all possible field configurations that have an imaginary time $\tau$ with period $\beta_\tau$. 
From the partition function, using standard thermodynamic relations one can obtain the Free Energy $\mathcal{G}$ of the system as 
\begin{equation}
    \mathcal{G} = - \frac{1}{\beta_\tau}\ln\mathcal{Z}~.
\end{equation}
By using the saddle point approximation (Laplace's method) we will consider that the classical action contributes the most and as a result we may drop the integral in the partition function $\mathcal{Z}$. 
Then the Euclidean action $\mathcal{I}_E$  can be related to the free energy evaluated on shell through the following relation
\begin{equation}
    \mathcal{I}_E = \mathcal{G} \beta_\tau ~.
\end{equation}
Having the expression of the free energy for the black hole solution, we will compare it with the free energy of the grand canonical ensemble in order to extract the mass (internal energy), the entropy, and the magnetostatic potential of the black hole. 
For more information in the discussion that follows, we refer the reader to the original work of Gibbons and Hawking \cite{Gibbons:1976ue}.

\subsection{GMGHS black hole}

We start our analysis with the thermodynamics of the GMGHS solution. The Euclidean action, including the appropriate boundary terms, is given by
\begin{equation}
    \label{euc-act-GMGHS}
    \mathcal I_E = - \frac{1}{16\pi}\int_\Sigma \diff^4x\sqrt{g}\left( \mathcal R-2\nabla_{\alpha}\phi\nabla^{\alpha}\phi -e^{-2\phi}\mathcal{F}^2\right) - \frac{1}{8\pi}\int_{\partial \Sigma}\left(K-K_0\right)d^3x\sqrt{h}~.
\end{equation}
In the Euclidean signature, the GMGHS black hole is described by the following metric: 
\begin{gather}
    \diff s^2 = B(r)\diff\tau^2 +\frac{\diff r^2}{B(r)} +[R(r)]^2 \diff\Omega^2~,
    %\hspace{1em}B(r)=\left(1-\frac{2M}{r}\right)\,,\hspace{1em}[R(r)]^2=r(r-Q_m^2/M)\,,\\[2mm]
   % \phi(r) = -\frac{1}{2} \ln\left(1-\frac{Q_m^2}{Mr}\right)~,\hspace{1em}
   % A_{\mu} = (0,0,0,Q_m \cos\theta)~.
    %\hspace{1em} F_{\mu\nu}=\partial_\mu A_\nu-\partial_\nu A_\mu\,.
\end{gather}
where $B(r)=1-2M/r$, while $R(r)$ has the same form as in eq. \eqref{B,R-r}.
In this coordinate system, the Euclidean time coordinate is periodic and takes values in the range $0\leq \tau \leq \beta_\tau$. For the derivation of the thermodynamic quantities, we assume that we have enclosed the black hole in a large cavity with radius $r_c$. Therefore, the radial coordinate takes values in $r_h \leq r < r_c$. Finally, the two angular coordinates take their usual values. The boundary term $K$ represents the trace of the extrinsic curvature, which in our case reads
\begin{equation}
    K = \nabla^{\alpha}n_{\alpha} =\frac{2R'(r)\sqrt{B(r)}}{R(r)} + \frac{B'(r)}{2\sqrt{B(r)}}\,,
\end{equation}
where $n_{\alpha} = \sqrt{1/B(r)}\,\delta^r_\alpha$ is a normalized spacelike vector field. The $K$ term in the above hypersurface integral represents the Gibbons-Hawking-York boundary term, ensuring a well-defined variational principle. The second boundary term $K_0$ serves as a subtraction term to render the action finite for flat space (in the absence of the black hole). For flat space, $K_0$ equals $2/r$, obtained by setting $B(r)=1$ and $R(r)=r$ in the above relation. Utilizing these relations, one can readily compute the Euclidean action \eqref{euc-act-GMGHS} to be
\begin{equation}
    \mathcal{I}_E = \frac{\beta_\tau  M}{2}-\frac{ \beta_\tau  Q_m^2}{4 M}~. \label{GHSeuclidean}
\end{equation}
In the above, we have used that the horizon radius is given by $r_h=2M$. 
In the Grand Canonical Ensemble, the Euclidean action is identified with the free energy of the thermodynamic system as $\mathcal{I}_E =\beta_\tau \mathcal{G}$, thus, we can rewrite (\ref{GHSeuclidean}) as
\begin{equation}
\label{free}
\mathcal{I}_E 
%\beta\tau M- \frac{\beta_\tau M}{2} - \frac{Q_m}{2M}Q_m\beta_\tau + \frac{Q_m^2}{4M}\beta_\tau =
%\beta_\tau M -\beta_\tau\Phi_m Q_m -4\pi M \left(M - \frac{Q_m^2}{2M}\right)
= \beta_\tau M -\beta_\tau\Phi_m Q_m -S~,
\end{equation}
where $S$ is the entropy and $\Phi_m$ is the magnetostatic potential, $\Phi_m=Q_m/r_h$.
For the derivation of the above equation, we have used the fact that $\beta_\tau=8\pi M \equiv 1/T$ with $T$ being the temperature of the black hole. 
By combining now eqs. \eqref{GHSeuclidean} and \eqref{free} we can evaluate the black-hole entropy $S$, which is given by the following relation
\begin{equation}
S = 2\pi M \left(2M - \frac{Q_m^2}{M}\right) =\pi [R(r_h)]^2= \frac{A}{4},\label{firstlaw}
\end{equation}
where $A$ denotes the horizon area. It is evident that in this case, the entropy function has the well-known form of the Bekenstein-Hawking entropy. For validation, the same result may also be obtained using Wald's formula or even using the Arnowitt-Deser-Misner (ADM) formalism \cite{ADM}.
For a comprehensive analysis of the ADM formalism, readers are directed to \cite{Corichi:1991qqo}. Additionally, for its explicit application in black-hole solutions, we refer the interested reader to \cite{Martinez:2004nb}.
In the subsequent subsection, we will utilize the ADM formalism for the thermodynamic analysis of our black-hole solution.

The inclusion of the Gibbons-Hawking-York boundary term ensures that the Euclidean action attains an extremum within the class of fields considered here, $\delta \mathcal I_{E} = 0~.$
As a result, it is evident that the first law of thermodynamics in the Grand Canonical Ensemble (keeping the temperature and the magnetic potential fixed) takes the form
\begin{equation}\label{firstlaw2}
\delta M = T \delta S + \Phi_m \delta Q_m~,
\end{equation}
derived from (\ref{free}), and holds \textit{by construction}. The first law is also evident by taking the variation of the entropy with respect to the primary black-hole charges.
The temperature of this black hole is the same as that of the Schwarzschild black hole, as pointed out in \cite{Horowitz:1992jp}, since the Euclidean continuation does not care about the angular part. Consequently, the heat capacity $C$ for constant charge will also be negative, $C=-1/(8 \pi T^2)$; hence, these types of black holes cannot reach thermal equilibrium.

\subsection{Black hole with non-linear electrodynamics}

We will now focus on our black-hole solution, emanating from the action \eqref{theory} and characterized by the line element \eqref{line-elem-phys}-\eqref{W-phys}.
The scalar and the gauge fields are of the form $\phi=\phi(R)$---with $R$ being the physical radial coordinate---and $\mathcal{A}_\mu=(0,0,0,A(\theta))$, respectively. 
In this case, to determine the thermodynamic quantities associated with the resulting black-hole solution, we will make use of the Euclidean signature and also utilize the ADM formalism \cite{ADM, Corichi:1991qqo}.
Hence, we consider the line element of the form
\begin{equation}
    \diff s^2 = [N(R)]^2B(R) \diff\tau^2+\frac{[W(R)]^2\diff R^2}{B(R)} + R^2 \diff\Omega^2~, \label{euclideanmetric}
\end{equation}
where the Euclidean time takes values in the range $0\le\tau\le\beta_\tau$, while the radial coordinate $R\in [R_h,+\infty)$.
To obtain the temperature, that is the period of the Matsubara frequency $\tau$, in our case, we follow the calculation of \cite{Karakasis:2023hni}. 
 To this end, we first ignore the angular part of the line element and perform a series expansion near the horizon.
Thus, we are left with a two-dimensional line element which is compared with the line element of two-dimensional space expressed in polar coordinates $\diff S=\diff\hat{R}^2 + \hat{R}^2\diff\Theta^2$.
By doing so, we obtain
\begin{eqnarray}
    &&\diff\hat{R}^2 = \frac{W(R_h)^2}{B'(R_h)(R-R_h)}\diff R^2~,\\[1mm]
    &&N^2(R_h)B'(R_h)(R-R_h)\diff\tau^2 = \hat{R}^2\diff \Theta^2~.
\end{eqnarray}
The coordinate $\Theta$ is periodic with a period $2\pi$ which implies that $\tau$ is also a periodic coordinate with a period $\beta_\tau$ given by:
\begin{equation}
    \beta_\tau = \frac{1}{T}= \frac{4\pi W(R)}{N(R)B'(R)}\Bigg|_{R_h}~,
    \label{beta}
\end{equation}
where $T$ is the temperature of the black hole.
For completeness, we also remark at this point that we have also checked that, as expected, the temperature will take on the same values at the event horizon regardless of the coordinate system we are using ($r$ or $R$).

The Euclidean action is related to the Lorentzian action via $\mathcal{I}_E = -i \mathcal{S}$ and we will consider the following variational problem which basically consists of the theory (\ref{theory}) alongside a boundary term denoted by $\mathcal{B}_E$ which we will consider in order to have a well-defined variational principle $\delta \mathcal{I}_E=0$.
Thus, we have
\begin{equation}
    \mathcal{I}_E = \frac{2\pi \beta_\tau}{16\pi} \int^{\pi}_{0}\diff\theta\int_{R_h}^{\infty}\diff R\left[-NR^2W\sin\theta\,\mathcal{L}(R,\theta)\right] +  \mathcal{B}_E~.
\end{equation}
Here $\mathcal{L}$ denotes the Lagrangian of the theory which is a function of $R,\theta$ coordinates.
After canceling total derivatives, the Euclidean action reads
\begin{align}
    \mathcal{I}_E =\beta_\tau \int^{\pi}_{0}\diff\theta\int_{R_h}^{\infty}\diff R\,\hat{\mathcal{L}}(Q^i,\partial_\mu Q^i)+ \mathcal{B}_E~\,,
\end{align}
with $Q^i=\{N(R),W(R),B(R),\phi(R),A(\theta)\}$ and $\hat{\mathcal{L}}(Q^i,\partial_\mu Q^i)$ given by
\begin{align}
    \hat{\mathcal{L}}(Q^i,\partial_\mu Q^i)=\frac{  N \sin \theta }{4 W^{2} R^6} \bigg[W R^7 B'&+W^{3} \left(2 f(\phi) (\alpha -\beta ) \frac{(\partial_\theta A)^4}{\sin^4\theta}+e^{-2\phi} R^4 \frac{(\partial_\theta A)^2}{\sin^2\theta} -R^6\right)  \nonumber\\ 
    &+B R^6 \left(W R^2 \phi'^2-2 R W'+W\right)\bigg]\,.
    \label{eq: ADM-L}
\end{align}
%
%\corr{($e^{-2\phi}=f_1(\phi)$)}.
Following the ADM formalism, we have to vary the above Euclidean action with respect to each one of the dynamical fields $Q^i$ to obtain the field equations.
By doing so, we obtain none other than the well-known Euler-Lagrange equations, namely
\begin{equation}
    \frac{\partial\hat{\mathcal{L}}}{\partial Q^i}-\partial_\mu\left(\frac{\partial\hat{\mathcal{L}}}{\partial(\partial_\mu Q^i)}\right) =0\,.
    \label{eq: E-L}
\end{equation}
Let us now apply the above equation for the dynamical field $Q^1=N(R)$.
Upon substituting the expression of $\hat{\mathcal{L}}$ from \eqref{eq: ADM-L} into \eqref{eq: E-L}, we find that the second term vanishes identically, while $\partial \hat{\mathcal{L}}/\partial N =\hat{\mathcal{L}}/N$.
As a result, the equation for $N(R)$ indicates that $\hat{\mathcal{L}}=0$, which in turn implies $\mathcal{I}_E = \mathcal{B}_E$.
This outcome is anticipated in the ADM formalism, where the metric construction \eqref{euclideanmetric} is specifically tailored to yield this result.
Additionally, by solving the field equations \eqref{eq: E-L} for all dynamical fields $Q^i$, one can determine the unknown functions and verify that the resulting solution is the one obtained in Sec. \ref{sec:model} with line element \eqref{line-elem-phys}-\eqref{W-phys}, alongside a constant $N$ which without loss of generality we may set equal to $1$.
It is important to mention at this point, that during the derivation of the Euler-Lagrange equations, certain boundary terms were omitted.
These terms are of the following form
\begin{equation}
    \beta_\tau\left(\frac{ R}{2 W}\delta B + \frac{2B   R^2  \phi'}{ W}\,\delta\phi- \frac{B R}{W^2}\,\delta W\right)\Bigg|_{R_h}^{\infty}~,
    \label{bound}
\end{equation}
and
\begin{equation}   
    \beta_\tau\int_{R_h}^{\infty}\diff R\left(\frac{  W e^{-2\phi} (\partial_\theta A)}{2 R^{2}\sin\theta}+ \frac{2 (\alpha-\beta)  W f(\phi) (\partial_\theta A)^3}{ R^{6}\sin^3\theta}\right)\delta A\Bigg|_{\theta=0}^{\theta=\pi}~. \label{magneticpot}
\end{equation}
The variation of the boundary term $\delta \mathcal{B}_E$ will account for the neglected boundary terms, ensuring the attainment of a well-defined variational principle $\delta \mathcal{I}_E=0$.
Utilizing the fact that the variation of $A$ yields $\delta A =(\delta Q_m) \cos\theta$, and substituting the expressions for the functions in (\ref{magneticpot}), one can integrate and derive the following expression:
\begin{equation}
    \beta_\tau\,\frac{ Q_m \left\{\sqrt{4 M^2 R_h^2+Q_m^4} \left[R_h^4-4  (\alpha-\beta ) Q_m^2\right]-R_h^4 Q_m^2\right\}}{2 M R_h^6}\,\delta Q_m~.
    \label{PhidQ}
\end{equation}
Now, the variation of the dynamical fields at infinity yield
\begin{eqnarray}
    &&\delta B = -\frac{2\delta M}{R} + \mathcal{O}\left(\frac{1}{R^2}\right)~,\\
    &&\delta \phi = \frac{ Q_m}{M R}\delta Q_m-\frac{ Q_m^2}{2 M^2 R}\delta M + \mathcal{O}\left(\frac{1}{R^3}\right)~,\\
    &&\delta W= \frac{ Q_m^4}{4 M^3 R^2}\delta M-\frac{ Q_m^3}{2 M^2 R^2}\delta Q_m + \mathcal{O}\left(\frac{1}{R^3}\right)~,
\end{eqnarray}
while at the horizon we have that
\begin{eqnarray}
    &&\delta B|_{R_h} = -B'(R_h)\delta R_h~,\\
    &&\delta \phi|_{R_h} = \delta\phi(R_h) -\phi'(R_h)\delta(R_h)~,\\
   &&\delta W|_{R_h} = \delta W(R_h) -W'(R_h)\delta(R_h)~.
\end{eqnarray}
Note that the parameters $\alpha,\beta$ are fixed by the theory and thus not allowed to vary, while $M$ and $Q_m$ are pure integration constants allowed in the variation. 

As previously mentioned, to ensure a well-defined variational procedure, it is desirable to have $\delta \mathcal{I}_E=0$. 
For clarity and convenience, we will partition the variation of the boundary term $\delta \mathcal{B}_E$ into two components: one at infinity and another at the event horizon, expressed as:
\begin{equation}
    \delta \mathcal{B}_E = \delta \mathcal{B}_E(\infty)+ \delta \mathcal{B}_E(R_h)~.
\end{equation}
Evaluating now (\ref{bound}) at infinity and considering the variation of the boundary term at infinity  we find that a zeroth order contribution survives, which according to the variations of the fields leads to
\begin{gather}
    \frac{\beta _{\tau }}{2} R  \left( \delta B  -2 \delta W  \right) + \delta \mathcal{B}_E(\infty)=0\Rightarrow\\[0.5mm]
    \delta \mathcal{B}_E(\infty) = \beta_\tau\, \delta M~.
\end{gather}
On the other hand, eq. (\ref{bound}) at the horizon, alongside the variation of the boundary term at the horizon and (\ref{PhidQ}) results in
\begin{equation}
    2\pi R_h \delta  R_h + \beta_\tau\,\frac{ Q_m \left\{\sqrt{4 M^2 R_h^2+Q_m^4} \left[R_h^4-4  (\alpha-\beta ) Q_m^2\right]-R_h^4 Q_m^2\right\}}{2 M R_h^6}\,\delta Q_m + \delta \mathcal{B}_E(R_h)=0~,
\end{equation}
which might be written equivalently as
\begin{eqnarray}
    \frac{\delta A}{4} + \beta_\tau\Phi_m \delta Q_m +  \delta \mathcal{B}_E(R_h)=0~,
\end{eqnarray}
where we have used the fact that the area of the black hole is given by $A = 4\pi R_h^2$ and we have defined the magnetic potential as 
\begin{equation}
    \Phi_m = \frac{ Q_m \left\{\sqrt{4 M^2 R_h^2+Q_m^4} \left[R_h^4-4  (\alpha-\beta ) Q_m^2\right]-R_h^4 Q_m^2\right\}}{2 M R_h^6}~.
\end{equation}
Considering now that we are dealing with the Grand Canonical Ensemble, we keep the temperature and the magnetic potential of the system fixed and as a result we can drop the variations to obtain
\begin{eqnarray}
    &&\mathcal{B}_E(\infty) = \beta_\tau M~,\\
    &&\mathcal{B}_E(R_h) = -\frac{A}{4} - \beta_\tau\Phi_m Q_m~.
\end{eqnarray}
Therefore, the value of the Euclidean action is given by
\begin{equation}
    \mathcal{I}_E = \beta_\tau M-\frac{A}{4} - \beta_\tau\Phi_m Q_m~,
\end{equation}
and since the Euclidean action is related to the free energy $\mathcal{G}$ of the system via $\mathcal{I}_E = \beta_\tau \mathcal{G} = \beta_\tau \mathcal{M} - S -  \beta_\tau\Phi_m Q_m$ we can identify, by comparison the conserved black hole mass and the entropy of the black hole as
\begin{eqnarray}
    &&\mathcal{M} = M~,\\
    &&S = A/4~.
\end{eqnarray}
Finally, the First Law of Thermodynamics \eqref{firstlaw2} holds \textit{by construction} as in the GMGHS black hole.

%%%%%%%%%%%%%%%%%%%%%%
\begin{figure}[t]
    \centering
    \begin{subfigure}[b]{0.48\textwidth}
    \includegraphics[width=1\textwidth]{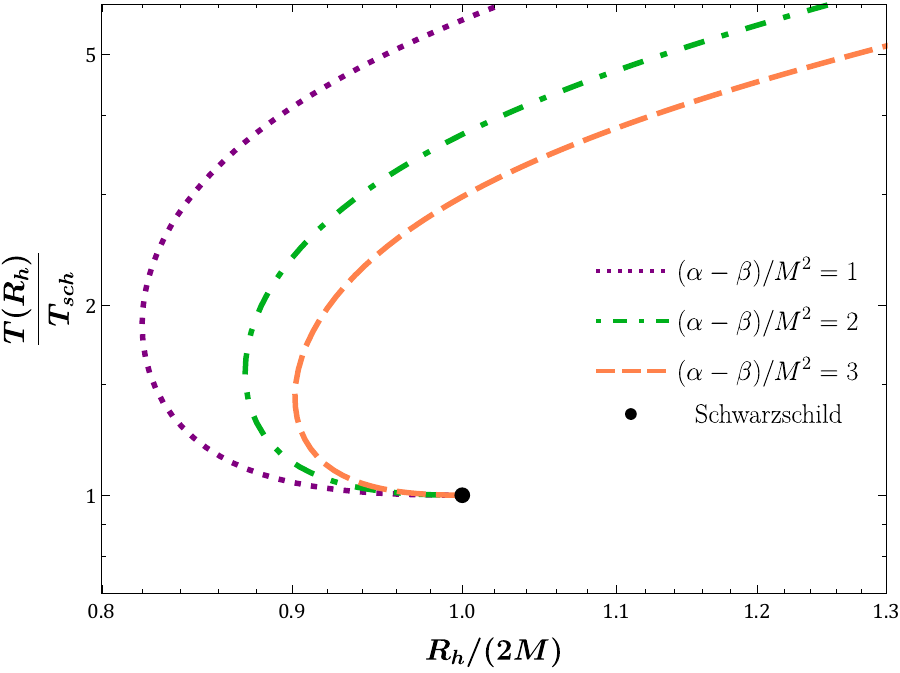}
    \caption{\hspace*{-2em}}
    \label{subf: T1}
    \end{subfigure}
    \hfill
    \begin{subfigure}[b]{0.492\textwidth}
    \includegraphics[width=1\textwidth]{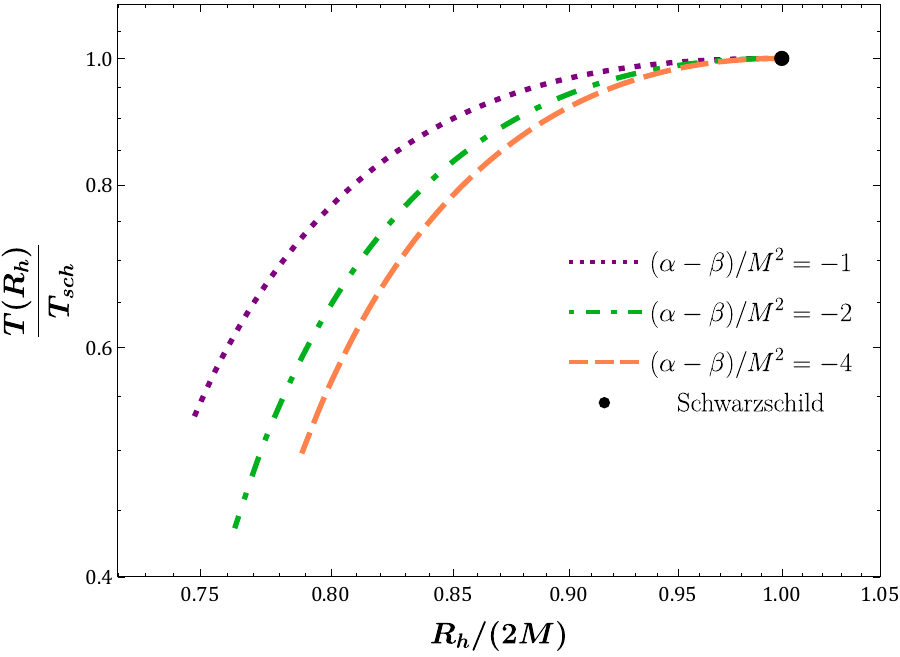}
    \caption{\hspace*{-2em}}
    \label{subf: T2}
    \end{subfigure}
    \caption{The black-hole temperature for (a) attractive and (b) repulsive higher-order electromagnetic contributions, with varying values of the magnetic charge ($Q_m$), while keeping the mass ($M$) the same. The axes in both figures are logarithmic.  }
    \label{fig: Temp}
\end{figure}
%%%%%%%%%%%%%%%%%%%%%%

With the confirmation that the black-hole thermodynamic quantities in our case adhere to the standard relations, we can now proceed to analyze the black hole's temperature.
In Fig. \ref{fig: Temp}, we depict the black-hole temperature as a function of the dimensionless quantity $R_h/(2M)$.
Notice that the temperature is scaled by the temperature of the Schwarzschild black-hole to form a dimensionless quantity, ensuring its independence from the chosen unit system.
In Figs. \ref{subf: T1} and \ref{subf: T2}, we explore the effects of the higher-order electromagnetic contributions on black-hole temperature, considering fixed (yet distinct) values for the coupling constants $\alpha$ and $\beta$, along with varying magnetic charge ($Q_m$) values, but maintaining the same value for the black-hole mass.
Both Figs. \ref{subf: T1} and \ref{subf: T2} were generated using the following procedure: 
For each value $(\alpha-\beta)/M^2$ and the ratio $Q_m/M$, we numerically evaluate the value of the ratio $R_h/(2M)$ using eq. \eqref{B-phys}.
Subsequently, employing equation \eqref{beta}, we calculate the temperature of the black hole for each parameter pair.
Finally, for each $(\alpha-\beta)/M^2$ we plot the points from the list $\{R_h/(2M),T(R_h)/T_{sch}\}$.
Note that we use the same mass parameter $M$ for the temperature calculation of the Schwarzschild black hole $T_{sch}$.
In both figures, we have also incorporated a distinctive dot symbolizing a constant value for the quantity $T(R_h)/T_{sch}$, irrespective of the ratio $R_h/(2M)$. 
Apparently, this is not coincidental, as it mirrors the characteristics of the Schwarzschild black hole, where the horizon radius precisely equals $2M$ and its temperature is determined by the established formula $T=1/(8\pi M)$.

Focusing now on the thermodynamical characteristics of our solution, we observe that regardless of the value $(\alpha-\beta)/M^2$, for $Q_m=0$, our solution reduces to the Schwarzschild black hole and therefore all graphs in Fig. \ref{fig: Temp} have as a starting point the Schwarzschild point.
However, when we depart from this limit, we notice that for $\alpha-\beta>0$, as illustrated in Fig. \ref{subf: T1}, the temperature of the resulting black holes consistently surpasses that of the Schwarzschild black hole, whereas for $\alpha-\beta<0$ (Fig. \ref{subf: T2}), the opposite effect occurs.
Furthermore, this temperature increase, in the $\alpha-\beta>0$ scenario, is independent of whether the black hole under examination possesses a smaller or larger horizon radius compared to the corresponding Schwarzschild black hole.
As previously observed in Fig. \ref{fig: Rh-Qm-M} and discussed in Sec. \ref{sec:model}, it becomes evident in Fig. \ref{subf: T1} that there are consistently pairs of black-hole solutions, more compact than the Schwarzschild solution, that share the same horizon radius $R_h$ but with different ratios $Q_m/M$.
However, now we see that although these solutions possess the same horizon radius, their temperatures differ significantly.
This can be understood through the relation \eqref{beta} where it is evident that the formula determining the temperature of a black hole depends on the first derivative of the function $B(R)$.
This means that black-hole solutions which for different ratios $Q_m/M$ result in the same horizon radius $R_h$ through the equation $B(R_h)=0$, their temperatures are not necessarily the same since $B'(R_h)$ could differ in these two cases.

Moreover, we can deduce the thermodynamic stability of these black holes by examining how the temperature changes with a change in the mass. In Fig. \ref{subf: T1} it is evident that there are two distinct branches of black-hole solutions. In the first branch we have black-hole solutions that get colder as the mass is decreasing, while in the second branch we have black holes that are getting hotter as the mass is getting smaller. This implies that the heat capacity $C \equiv dM/dT$ for the first branch is positive since both $dM,dT$ are negative and the black holes are thermally stable, while the second branch, for which the temperature rises with the decrease of mass, exhibits negative heat capacity and are thermally unstable. Notice also the fact that the Schwarzchild black hole lies in the second (unstable) branch which is a well-known result. Furthermore, the parameter space of these black holes exhibits a point where $dT=0$ indicating the divergence of the heat capacity and as a result a phase transition from hot to cold black holes. In Fig. \ref{subf: T2}, we can see that as the black holes lose mass they get colder which implies that they are thermally stable since they possess positive heat capacity. These results are in agreement with the studies in 
\cite{Chatzifotis:2023ioc}, where the black holes are viewed as defects in the thermodynamical spacetime~\cite{Wei:2022dzw}.  

In the next section, we proceed to study the stability of the black holes from a linear-perturbation point of view, which, in general, is distinct from the thermodynamic stability. Indeed, as we shall demonstrate explicitly below, such a linearised stability analysis does not necessarily imply thermodynamical stability, in the sense that the thermodynamically unstable branches found above exhibit stability under linear perturbations.

\section{Linear Perturbations}\label{sec:linear}

\subsection{Radial Stability}

In this section, we investigate the stability of our solution under radial perturbations. For simplicity, we focus on linear and radial perturbations. Therefore, we use the following ansatz:
\begin{equation}
    ds^2=-P(R,t)dt^2+Q(R,t)dR^2+R^2d\Omega^2, \qquad A=a_0(R,t)dt + Q_m \cos \theta\,d\varphi, \qquad \phi=\phi(R,t),
\end{equation}
%%%%
where
%%%%%
\begin{gather}
    P(R,t)=B(R)\left[1-\epsilon\, e^{-i\omega t} h_1(R) \right], \qquad Q(R,t)=\frac{1}{H(R)}+\epsilon\, e^{-i\omega t} h_2(R), \\[2mm]
    \phi(R,t) = \phi(R)+\epsilon\, e^{-i\omega t} \phi_1(R), \qquad a_0(R,t)=\epsilon\, e^{-i\omega t} a_0(R).
\end{gather}
For the stability analysis, it is more convenient to work within the physical coordinate system; hence, in the above equations $B(R)$, $\phi(R)$ and $H(R)=B(R)/W^2(R)$ are given in eqs. (\ref{B-phys}--\ref{phiphys}), and correspond to the background/unperturbed spacetime. Note that the radial perturbations are associated with the $L=0$ perturbation\,\footnote{$L$ is the ``angular momentum" index in the spherical harmonics function $Y_L^{M_L}(\theta,\varphi)$. For more information about the decomposition of the perturbations in spherical harmonics, see \cite{Regge:1957td,Zerilli:1970se,Zerilli:1970wzz,Zerilli:1974ai}.} in the even sector of the gravitational perturbations. Therefore, in the electromagnetic part, only the electric-type perturbations contribute, as the magnetic-type corresponds to the odd sector \cite{Zerilli:1974ai, Astefanesei:2019qsg}. The dimensionless constant $\epsilon$ determines the order of the perturbation. Finally, $\omega$ specifies the decomposition of the modes with fixed energy. 

%%%%%%%%%%%%%%%%%%%%%%
\begin{figure}[ht]
    \centering
    \begin{subfigure}[b]{0.49\textwidth}
    \includegraphics[width=1\textwidth]{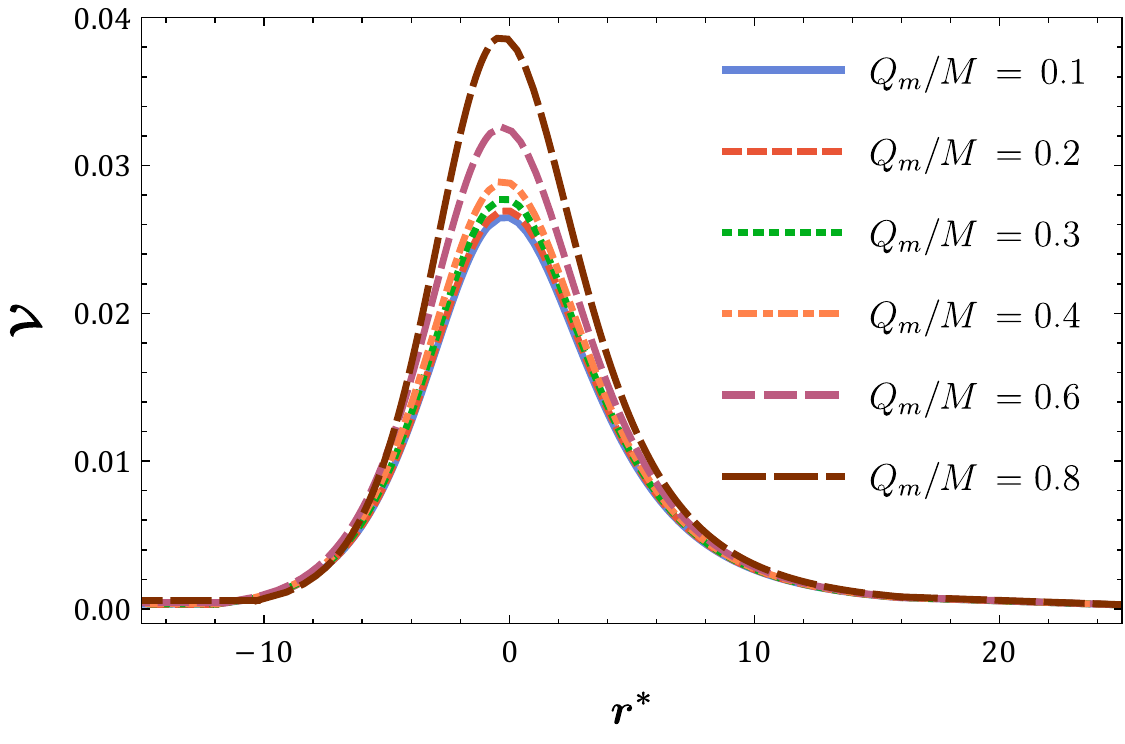}
    \caption{\hspace*{-3em}}
    \label{subf: spotpos}
    \end{subfigure}
    \hfill
    \begin{subfigure}[b]{0.48\textwidth}
    \includegraphics[width=1\textwidth]{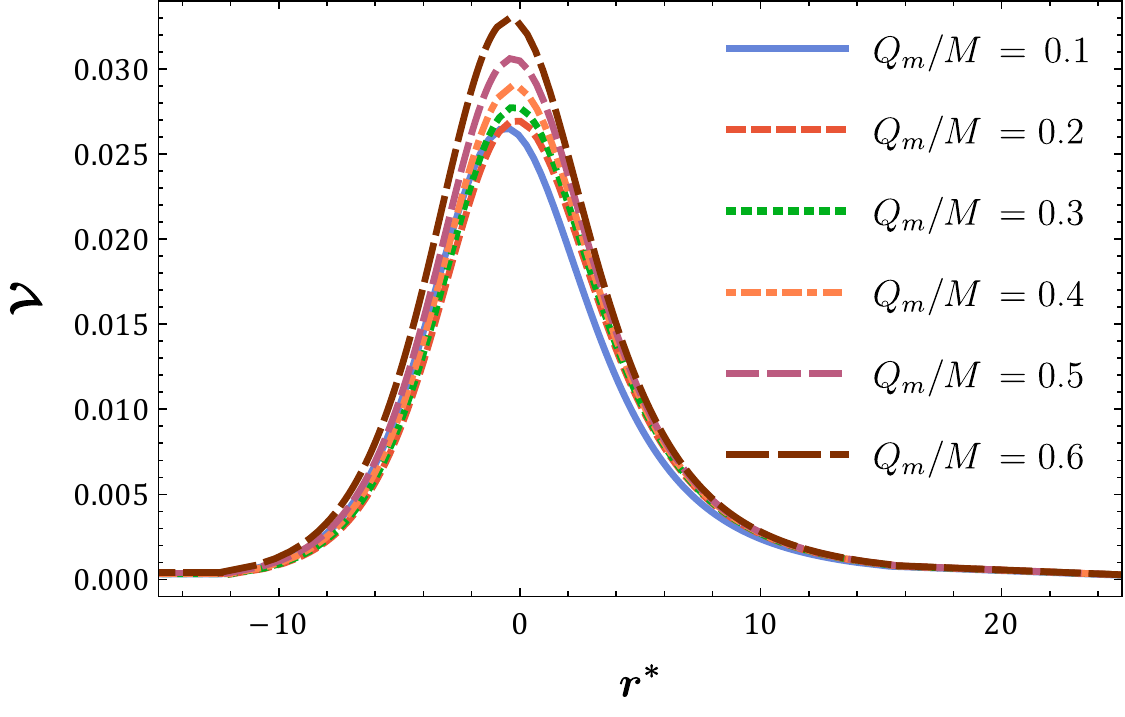}
    \caption{\hspace*{-3.3em}}
    \label{subf: spotneg}
    \end{subfigure}
    \caption{The graph of the effective potential $\mathcal{V}$ in terms of $r^*$ for various values of the parameter $Q_m/M$ and (a) $(\alpha-\beta)/M^2=1$, (b) $(\alpha-\beta)/M^2=-1$.}
    \label{fig: spot}
\end{figure}
%%%%%%%%%%%%%%%%%%%%%%

A direct calculation reveals that both the spacetime and the matter field perturbations are determined from the function $\phi_1$. Consequently, the investigation of the system is simplified to a single equation for the perturbation of the scalar field.  This specific equation can be expressed in the conventional Schrödinger form:
%%%%%
\begin{equation}\label{schr}
    \frac{d^2\Psi(r^*)}{{d r^*}^2}+ \left[ \omega^2 -\mathcal{V}(R)  \right]\Psi(r^*)=0,
\end{equation}
%%%%%%%
where we have defined a new perturbation function as $\Psi\equiv R \phi_1$ and the tortoise coordinate is $dr^*\equiv\frac{dR}{\sqrt{B(R)H(R)}}$. The potential of the Schrödinger equation has the following form
\begin{align}\label{schpotr}
    \mathcal{V}(R)=&\,\frac{H B'+2 B R \phi ' \left[\phi ' \left(R \left(H'+H R \phi
   '^2\right)+4 H\right)+3 H R \phi ''\right]+B H'}{2 R} \nonumber\\[1mm]
   &+\frac{B (\alpha
   -\beta ) Q_m^4 \left(R \dot f \phi '+\ddot{f}\right)}{R^8}+\frac{B \gamma  e^{-2
   \phi } Q_m^2 \left(2-R \phi '\right)}{R^4}\,.
\end{align}
%%%%%%
Considering our emphasis on the stability of black-hole solutions, there is no need to solve eq. (\ref{schr}).
The time evolution factor, $\exp{(-i\omega t)}$, simplifies the task, requiring us only to ascertain whether the frequency, $\omega$, is purely imaginary or not. 
In the scenario where the frequency $\omega$ is purely imaginary,  the mode undergoes exponential growth due to the presence of the term $\exp{(-i\omega t)}$ making the solution unstable. 
Therefore,  a negative eigenvalue, $\omega^2<0$, that signifies an unstable mode,  corresponds to a bound state in the Schrödinger equation (\ref{schr}). 
A general result in quantum physics is that for a potential that vanishes in both asymptotic regions, has a barrier form, and is positive definite.
Therefore, eq. (\ref{schr}) does not exhibit bound states. In Fig. \ref{fig: spot}, we depict the potential of the Schrödinger equation for two families of solutions. The first one corresponds to $(\alpha-\beta)/M^2=1$, while the second one corresponds to $(\alpha-\beta)/M^2=-1$. By carefully examining the parametric space of the solutions, we deduce that the potential always takes a form similar to the potentials in Fig. \ref{fig: spot}. 
Therefore, we conclude that our solutions are stable under radial perturbations. 

Although radial stability is a strong indication regarding the stability of a particular solution, a more careful and general perturbation analysis has to be performed to extract a stronger result, which, however, lies beyond the scope of this work. Moreover, as shown in the previous section \ref{sec:thermo}, linear stability does not necessarily imply thermodynamical stability for the black hole, in the sense that the latter, although linearly stable, nonetheless it possesses thermodynamically unstable branches.

\subsection{Scalar Quasi-Normal Modes}

Quasi-normal modes (QNMs) play a crucial role in the study of black holes and other astrophysical objects \cite{Kokkotas:1999bd, Berti:2009kk, Konoplya:2011qq}. These modes represent the characteristic vibrations or oscillations of a black hole after a perturbation, such as a gravitational wave or a scattering event. QNMs are characterized by complex frequencies, i.e. eigenvalues of the Schrödinger equation, consisting of a real part and an imaginary part. The real part corresponds to the oscillation frequency, while the imaginary part reflects the damping or decay of the mode. The study of QNMs provides valuable insights into the nature and properties of black holes, offering a unique window into their internal dynamics. 

For simplicity, we will consider the propagation of a test scalar field $\Phi$ in the background of our black hole and extract its QNMs. We begin  our analysis from the following action functional for the scalar field, 
%%%
\begin{equation}
    S=\int d^4x \sqrt{-g}\left[\nabla^\mu\Phi \nabla_\mu \Phi + m^2 \Phi^2 \right],
\end{equation}
%%%%%
where $m$ is the mass of the test scalar field. The variation of the above action with respect to the scalar field  yields the Klein-Gordon equation in the black hole background 
%%%%%
\begin{equation}
    \left(\square - m^2\right)\Phi=0\,.
\end{equation}
%%%%%%%
Note that the test scalar field $\Phi$, as a perturbation field, does not back-react on the spacetime metric and is a function of all spacetime coordinates $\Phi = \Phi(t,R,\theta,\varphi)$. 
For clarity, we choose to work in the physical coordinate system. Therefore, the background metric is given by eq. (\ref{line-elem-phys}).  We can apply the separation of variables as follows
%%%%%%%
\begin{equation}
\Phi(t,R,\theta,\varphi) = e^{-i\omega t}Y_L^{M_L}(\theta,\varphi)\frac{\Uppsi(R)}{R}~,
\end{equation}
%%%%%%%%
where $Y_L^{M_L}(\theta,\varphi)$ represents the spherical harmonics function. By using the tortoise coordinate, $dr^*=\frac{dR}{\sqrt{B(R)H(R)}}$, one can rewrite the perturbation equation in a Schrödinger form as:
%%%%%
\begin{equation}\label{schq}
    \frac{d^2\Uppsi(r^*)}{{d r^*}^2}+ \left[ \omega^2 -V(R)  \right]\Uppsi(r^*)=0,
\end{equation}
%%%%%%%
where $V(R)$ is the effective potential of the Schrödinger equation and is given by
%%%%%%%%
\begin{equation}
    V= \frac{HB'+BH'}{2R}+\frac{B}{R^2}L(L+1)+m^2B.
\end{equation}
The background metric functions $B$ and $H=B/W^2$ are given in eqs. (\ref{B-phys}--\ref{W-phys}).  

In the pursuit of calculating the QNMs, the WKB (Wentzel-Kramers-Brillouin) approximation stands as a valuable method. Particularly useful in the context of wave-like phenomena, the WKB approximation provides an efficient and semi-classical approach to estimating the complex frequencies associated with QNMs. By treating the Schrödinger-like equation governing the perturbations as a semiclassical wave equation, the WKB method allows for the determination of QNM frequencies without the need for an exact solution. The WKB method was initially developed for quantum mechanical problems; however, Schutz and Will were the first to apply this method to the problem of scattering around black holes \cite{Schutz:1985km}. Later, Iyer and Will extended this approach to the third WKB order beyond the eikonal approximation \cite{Iyer:1986np}, and Konoplya further advanced it to the sixth order \cite{Konoplya:2003dd,Konoplya:2011qq}. Interestingly, the sixth order yields a relative error of approximately two orders of magnitude lower than that of the third WKB order \cite{Konoplya:2003dd,Konoplya:2011qq}. However, for simplicity, in this work, we will employ the first-order WKB approximation, in which the QNM frequencies are obtained from the solution of the following equation 
%%%%
\begin{equation}
    n+\frac{1}{2}=-i \left[\frac{\omega^2-V(r^*)}{\sqrt{-2V''(r^*)}}\right]_{r*=r^*_{\rm max}},\qquad n\in \mathbb{Z}^{\geq}.
\end{equation}
%%%%%
The expression in the right-hand part of the above equation is evaluated at the maximum of the potential $r^*_{\rm max}$ while $n$ is the overtone number of the QNMs.

\begin{table}[t]
\center{
\begin{tabular}{|c|c|c|c|}
 \hline
  
  & $(\alpha-\beta)/M^2=0.01$  & $(\alpha-\beta)/M^2=-1$ &  $(\alpha-\beta)/M^2=1$ \\
 \hline\hline
  $Q_m/M=0.01$ & $0.329438 - 0.096255 \,i$  & $0.329438 - 0.096255 \,i$   & $0.329438 - 0.096255 \,i$ \\
 \hline
   $Q_m/M=0.3\,\,$ & $0.333767 - 0.096712 \,i$  & $0.333767 - 0.096728 \,i$  & $0.333767 - 0.096696 \,i$  \\
 \hline
   $Q_m/M=0.6\,\,$ & $0.348235 - 0.098171 \,i$  & $0.348226 - 0.098595 \,i$ & $0.348243 - 0.097727 \,i$    \\
 \hline
\end{tabular}\label{QNM1}
\caption{The  dimensionless $L=1$, $n=0$ quasinormal modes  ($M \omega$) for $m=0$.}
\label{qnm-m0}
}
\end{table}

\begin{table}[t]
\center{
\begin{tabular}{|c|c|c|c|}
 \hline
  
  & $(\alpha-\beta)/M^2=0.01$  & $(\alpha-\beta)/M^2=-1$ &  $(\alpha-\beta)/M^2=1$ \\
 \hline\hline
  $Q_m/M=0.01$ & $0.401632 - 0.050195 \, i$  & $0.401632 - 0.050195 \,i$   & $0.401632 - 0.050195 \,i$  \\
 \hline
   $Q_m/M=0.3\,\,$ & $0.404334 - 0.052296 \,i$  & $0.404335 - 0.052294 \,i$  & $0.404332 - 0.052298 \,i$ \\
 \hline
   $Q_m/M=0.6\,\,$ & $0.413700 - 0.058602 \, i$  & $0.413733 - 0.058522 \,i$  & $0.413667 - 0.058679 \,i$     \\
 \hline
\end{tabular}\label{QNM2}
\caption{The  $L=1$, $n=0$ dimensionless quasinormal modes ($M \omega$)   for  $m/M=0.4$.}
\label{qnm-m}}
\end{table}

In Tables \ref{qnm-m0} and \ref{qnm-m}, we present the dimensionless QNM frequencies, denoted by $(M\omega)$, for two distinct scenarios: when $m=0$ and $m/M=0.4$, respectively. Notably, as $Q_m$ approaches 0, our solution converges to the Schwarzschild black hole, irrespective of the $(\alpha-\beta)/M^2$ parameter. This convergence is evident in the first row of both tables, where $Q_m/M=0.01$, as the QNM values remain constant across varying $(\alpha-\beta)/M^2$. Furthermore, as $(\alpha-\beta)/M^2$ tends toward 0, our solution adopts the characteristics of the GMGHS black hole. Consequently, the QNMs in the first row of both tables, specifically when $(\alpha-\beta)/M^2=0.01$, align with those of the GMGHS black hole.
The subsequent rows in the tables provide additional insights into the characteristics of our solution. For instance, in the second and third rows, where $Q_m/M=0.3$ and $0.6$, respectively, we observe a systematic variation in the QNM values with changes in both $Q_m/M$ and $(\alpha-\beta)/M^2$. This behavior highlights the sensitivity of the QNM frequencies to the parameters characterizing the black hole solution. Furthermore, by comparing these results to the Schwarzschild and GMGHS cases, we discern how our solution deviates from these benchmark scenarios.
Additionally, the tables reveal intriguing patterns in the imaginary parts of the QNMs. For varying $Q_m/M$ and $(\alpha-\beta)/M^2$, the imaginary parts exhibit non-trivial changes, reflecting the impact of the black hole's charge and the parameter $(\alpha-\beta)/M^2$ on the damping behavior of the perturbations.

\section{Other Solutions}\label{sec:othersol}

\subsection{Asymptotically (A)dS spacetimes}

Let us now, briefly discuss asymptotically (A)dS spacetimes. Following \cite{Gao:2004tu}, introducing a scalar potential $\mathfrak{V}(\phi)$ in the action and considering
\begin{equation}
    \mathcal{S} = \int \diff^4x \sqrt{-g} \Big(\mathcal{R} - 2\nabla^{\mu}\phi\nabla_{\mu}\phi -e^{-2\phi}\mathcal{F}^2  +f(\phi)\big(   -2\alpha\mathcal{F}^{\alpha}_{~\beta}\mathcal{F}^{\beta}_{~\gamma}\mathcal{F}^{\gamma}_{~\delta}\mathcal{F}^{\delta}_{~\alpha}+\beta \mathcal{F}^4\big) - \mathfrak{V}(\phi) \Big)~,
\end{equation}
with a $\mathfrak{V}(\phi)$ of the form
\begin{equation}
    \mathfrak{V}(\phi) = \frac{1}{3} \Lambda  e^{-2 \phi }+\frac{1}{3} \Lambda  e^{2 \phi }+\frac{4 \Lambda }{3} = \frac{2}{3} \Lambda  (\cosh (2 \phi )+2)~,
\end{equation}
we can obtain $B(r)$ as
\begin{equation}
    B(r) = 1 -\frac{2 M}{r} - \frac{2  (\alpha -\beta ) Q_m^4}{r^3 \left(r-\frac{Q_m^2}{M}\right){}^3}-\frac{1}{3} \Lambda 
   r \left(r-\frac{Q_m^2}{M}\right)~,
\end{equation}
while $\phi(r), R(r)$ will remain the same. Note here that the potentials $\mathfrak{V}(\phi)$ and $f(\phi)$ are almost identical, and they are both Liouville-type potentials \cite{Charmousis:2009xr}.\footnote{Nonetheless, we should notice that the presence of a positive cosmological constant term in the gravitational effective actions is problematic within the context of string/brane-inspired quantum gravity theories, due to the so-called swampland conjecture~\cite{Obied:2018sgi,Palti:2019pca,Ooguri:2018wrx}. This issue is open at present, and its study goes beyond the purpose of the current article.}

\subsection{Solutions for general $\gamma$}

Assuming that the coupling term between the dilaton and the Maxwell term is of the form $e^{-2\gamma\phi}$ we can obtain the same geometry with the $\gamma=1$ case with the coupling function $f(\phi)$ now being given by
\begin{equation}
    f(\phi) =-3 \cosh (2 \phi )-2 -\frac{e^{2 \phi } \left(\left(e^{-2 \phi }\right)^{\gamma -1}-1\right) Q_m^6}{2 M^4
   \left(e^{2 \phi }-1\right)^4 (\alpha -\beta )}~.
\end{equation}
In this case, the charge-to-mass ratio is fixed by the theory. As a result, such black holes are described by a constrained phase space of free parameters, since this situation reduces the number of primary black hole hairs from two to one. A more physical result would be to let the form of the dilaton field to be affected by the change of the coupling function with the Maxwell term, however, we were not able to derive exact results in this case, so one has to employ numerical techniques. Such endeavors may be undertaken in subsequent works.

\section{Conclusions}\label{sec:concl}

In the quest to comprehend gravitational phenomena and the nature of gravity itself, the theoretical exploration of black holes stands as a pivotal frontier. The predictions of General Relativity (GR) are in good agreement with current observations related to black holes. This is attributed to the large mass of the observed objects and therefore their large horizon radius and small horizon curvature.  Additionally, a plethora of cosmological observations indicates instances where GR exhibits limitations, with the most notable challenges being the Dark Energy Problem and GR’s inability to account for the inflationary epoch in our universe. Therefore, the validity of General Relativity is expected to come under scrutiny in extreme conditions. General Relativity is commonly acknowledged as an effective theory applicable only within the realm of low energies. Consequently, such observations motivate us to explore modified gravitational theories, especially in extreme conditions where GR's validity may be compromised. Among these theoretical frameworks, modifications originating from String Theory, particularly the heterotic string theory, emerge as leading contenders. Notably, String Theory offers insights into high-order corrections, ranging from the Gauss-Bonnet term to non-linear electromagnetic effects, and provides a rich avenue for exploring the behavior of black holes under diverse conditions.

One intriguing aspect of string/brane-induced non-linear electrodynamics is the emergence of the Born-Infeld (BI) Lagrangian, which encapsulates higher-order corrections to Maxwell's theory. This Lagrangian arises from the resummation of open string excitations, particularly in the context of D-brane worlds in string theory. The coupling of the BI Lagrangian to the dilaton field in curved spacetime leads to an effective four-dimensional action, offering a novel perspective on electromagnetic interactions in the presence of gravity. Furthermore, considerations of higher-order electromagnetic terms, originating from closed string sectors, broaden the theoretical landscape. The inclusion of string loops leads to generalized effective actions, incorporating both closed and open string contributions, and potentially revealing novel phenomena beyond conventional electromagnetic frameworks. Departing from traditional electromagnetic theories, the exploration of non-linear electrodynamics within the context of black hole solutions offers a rich avenue for understanding strong-field regimes and cosmological implications. Non-linear effects become crucial in regions with intense gravitational fields, such as those near black holes, shedding light on phenomena absent in linear theories. Moreover, non-linear electrodynamics holds relevance for early universe cosmology, where the interplay between gravitational and electromagnetic fields played a significant role. 

In this work, we considered a string-inspired theory that involves a scalar field $\phi$ coupled to the electromagnetic field via a non-linear function $f(\phi)$. The action encompasses higher-order electromagnetic invariants, contributing to the field equations and leading to novel black hole solutions. Furthermore, we investigated the impact of a non-trivial coupling function $f(\phi)$, considering a specific functional form motivated by string-inspired models. The resulting exact, magnetically charged black hole solution revealed significant departures from the classical General Relativity predictions, with the scalar field and the electromagnetic field configurations exhibiting non-trivial behavior. We explored the implications of different coupling constants $\alpha$ and $\beta$ on the spacetime geometry and electromagnetic field configurations. The solutions obtained exhibit intriguing features, including dependence on the sign of $\alpha-\beta$ which determines whether the higher-order electromagnetic terms contribute attractively or repulsively to the spacetime geometry. Additionally, we examined the horizon structure of the black hole solutions, observing transitions from single to multiple horizons and even to naked singularities as the parameters varied. Notably, the compactness of the black holes relative to the Schwarzschild solution depended on the the magnetic charge to mass ratio. Our findings suggest a rich interplay between the scalar field, electromagnetic field, and spacetime geometry, highlighting the potential implications of such theories in astrophysical contexts, and the search for potential signatures of string theory in black hole physics, at least those signatures that can be manifested through effective string-inspired field theory models. It goes without saying, however, that the present work does not deal with a detailed experimental sensitivity analysis of such objects, which still remains to be done.

The examination of geodesics and energy conditions delves into the intricate dynamics of particles moving within the spacetime geometry described by the black hole solutions under investigation. By analyzing the geodesic equations we unveil the behavior of massive particles in the vicinity of these black holes, elucidating the role of the effective gravitational potential. Notably, the effective potential exhibits distinct features depending on the relative values of the coupling constants, offering insights into the stability and nature of orbits around the black holes. Additionally, the examination of energy conditions associated with the effective stress-energy tensor reveals intriguing properties of the spacetime, indicating the existence of a dilaton hair in the black hole's exterior while satisfying all energy conditions. This observation challenges the traditional no-hair theorems, underscoring the nuanced interplay between gravitational theories, non-linear electrodynamics, and scalar fields in modified theories. Our thermodynamic analysis provided valuable insights into the properties of black holes in both the GMGHS solution and our black-hole solution with non-linear electrodynamics. Our analysis allows for the extraction of important thermodynamic quantities such as mass, entropy, magnetostatic potential, and the extraction of the first Law of Thermodynamics. Notably, the entropy of both black hole is consistent with the Bekenstein-Hawking entropy formula. By examining the behavior of the temperature, we concluded that when the non-linear electrodynamics terms act attractively, there exist two distinct branches of black holes, one that is getting colder as the mass is decreasing and therefore is thermally stable, and another one that is getting hotter as the mass is decreasing which is thermally unstable. On the other hand, when the non-linear electrodynamics terms have a repulsive effect, the black holes are getting colder as the mass is decreasing and as a result are thermally stable.

Finally, our analysis of linear perturbations and scalar quasi-normal modes (QNMs) provides valuable insights into the stability and dynamic behavior of black hole solutions with non-linear electrodynamics. Through a rigorous investigation of radial stability, we demonstrated that our black hole solutions remain stable under linear and radial perturbations. This finding underscores the robustness of our black hole solutions against radial perturbations, supporting their viability as physically meaningful configurations within the framework of non-linear electrodynamics. Furthermore, our examination of scalar QNMs yielded intriguing results regarding the characteristic vibrations and oscillations of the black hole spacetime. Moreover, the analysis of scalar QNMs revealed the intricate interplay between the black hole parameters, such as charge and $(\alpha-\beta)/M^2$, and the frequency and damping behavior of perturbations. By systematically varying these parameters, we observed distinct patterns in the QNM frequencies, indicating the sensitivity of the black hole's dynamic properties to its intrinsic characteristics. Notably, our results exhibited convergence to the Schwarzschild black hole in the limit of vanishing charge and alignment with the GMGHS black hole in specific parameter regimes. These observations 
%emphasize the nuanced influence of charge and the $(\alpha-\beta)/M^2$ parameter on the black hole's dynamic behavior and stability, highlighting
highlight the rich phenomenology associated with black holes in our string-inspired theory.

\section*{Acknowledgements}

This work was supported by the Hellenic Foundation for Research and Innovation (H.F.R.I.) under the “3rd Call for H.F.R.I. Research Projects to support Post-Doctoral Researchers” (Project Number: 7212).
The work of NEM was supported in part by the UK Science and Technology Facilities research
Council (STFC) and UK Engineering and Physical Sciences Research Council (EPSRC) under the research grants  ST/X000753/1 and  EP/V002821/1, respectively.
NEM also acknowledges participation in the COST Association Action CA21136 “Addressing observational tensions in cosmology with systematics and fundamental physics (CosmoVerse)”.

\appendix

\bibliography{Refs}{}
\bibliographystyle{utphys}

\end{document}